\PassOptionsToPackage{table,xcdraw}{xcolor}
\documentclass[12pt,a4paper]{article}
\pdfoutput=1
\usepackage{jheppub}
\usepackage{gensymb}
\DeclareSymbolFont{matha}{OML}{txmi}{m}{it}% txfonts
\DeclareMathSymbol{\varv}{\mathord}{matha}{118}
\usepackage{amsmath}
\usepackage{amssymb}
\usepackage{graphicx}
\usepackage{subfigure}
\usepackage{pstricks}
\usepackage{bm}
\usepackage{pbox}
\usepackage{placeins}
\usepackage{booktabs}
\usepackage[T1]{fontenc}
\usepackage{footnote}
\usepackage{pdfpages}
\usepackage{hhline}
\usepackage{multirow}
\usepackage{multicol}
\usepackage{enumitem}
\usepackage[toc,page]{appendix}
\allowdisplaybreaks
\usepackage{comment}
\usepackage{float}
\usepackage{slashed}
\usepackage{xcolor}
\usepackage{textcomp}
\usepackage{gensymb}
\usepackage{multirow}
\usepackage[numbers]{natbib}
\usepackage{notoccite}
\usepackage{soul}
\usepackage{rotating}
\usepackage{tabularx}
\usepackage[labelfont=bf]{caption} % optional

\FloatBarrier

\usepackage[many]{tcolorbox}

\renewcommand{\thetable}{\Roman{table}}

\renewcommand{\thetable}{\Roman{table}}
\newcommand{\matrixel}[3]{\left< #1 \vphantom{#2#3} \right|
 #2 \left| #3 \vphantom{#1#2} \right>} 

\definecolor{MyDarkBlue}{rgb}{0.1, 0.1, 0.8} 
\definecolor{MyLightBlue}{rgb}{0.22,0.51,0.9}
\definecolor{MyGreen}{rgb}{0.0, 0.5, 0.0}
\definecolor{BrickRed}{rgb}{0.8, 0.25, 0.33}

\RequirePackage{hyperref}
\hypersetup{colorlinks, citecolor=BrickRed,linkcolor=blue, urlcolor=magenta}

\makeatletter
\gdef\@fpheader{}
\makeatother
\begin{document}

\title{\Large
Gauge and Scalar Boson Mediated Proton Decay in a Predictive SU(5) GUT Model
}
\author[a,b]{Ilja Dor\v{s}ner,}
\author[c,d]{Emina D\v{z}aferovi\'c-Ma\v{s}i\'c,}
\author[b,e]{Svjetlana Fajfer,}
\author[f]{and Shaikh Saad}

\affiliation[a]{University of Split, Faculty of Electrical Engineering, Mechanical Engineering and Naval Architecture in Split, Ru\dj era Bo\v skovi\' ca 32, HR-21000 Split, Croatia}
\affiliation[b]{J.\ Stefan Institute, Jamova 39, P.\ O.\ Box 3000, SI-1001 Ljubljana, Slovenia}
\affiliation[c]{University of Zagreb, Faculty of Science, Department of Physics, Bijeni\v{c}ka cesta 32, HR-10000 Zagreb, Croatia}
\affiliation[d]{University of Sarajevo, Faculty of Mechanical Engineering, Vilsonovo \v setali\v ste 9, BA-71000 Sarajevo, Bosnia and Herzegovina}
\affiliation[e]{Department of Physics, University of Ljubljana, Jadranska 19, SI-1000 Ljubljana, Slovenia}
\affiliation[f]{Department of Physics, University of Basel, Klingelbergstrasse\ 82, CH-4056 Basel, Switzerland}

\emailAdd{dorsner@fesb.hr,  emina.dzaferovic@mef.unsa.ba, svjetlana.fajfer@ijs.si, shaikh.saad@unibas.ch}
%%%%%%%%%%%%%%%%%%%%%%%%%%%%%
\abstract{
We assess proton decay signatures in the simplest viable $SU(5)$ model with regard to constraints on parameters governing the Standard Model fermion mass spectrum. Experimental signals for all eight two-body proton decay processes result from exchange of two gauge bosons, a single scalar leptoquark, or their combination. Consequently, it enables us to delve into an in-depth anatomy of proton decay modes and anticipate future signatures. Our findings dictate that observing a proton decay into $p\to\pi^0e^+$ indicates gauge boson mediation, with the potential for observation of $p\to\eta^0e^+$ mode. Alternatively, if decay is through $p\to K^+\overline\nu$ process, it is mediated by a scalar leptoquark, possibly allowing the observation of $p\to\pi^0\mu^+$. Detection of both $p\to\pi^0 e^+$ and $p\to K^+\overline\nu$ could enhance $p\to\pi^0\mu^+$ through constructive interference. The model predicts inaccessibility of $p\to\pi^+\overline\nu$, $p\to\eta^0\mu^+$, $p\to K^0e^+$, and $p\to K^0\mu^+$, regardless of the dominant mediation type, in the coming decades. In summary, through a comprehensive analysis of proton decay signals, gauge coupling unification, and fermion masses and mixing, we both accurately and precisely constrain the parameter space of the $SU(5)$ model in question.
}
\maketitle

%%%%%%%%%%%%%%%%%%%%%%%%%%%%%%%%%%%%%%%%%%%%%%%
%%%%%%%%%%%%%%%%%%%%%%%%%%%%%%%%%%%%%%%%%%%%%%%
\section{Introduction}

Grand Unified Theories~\cite{Pati:1973rp,Pati:1974yy, Georgi:1974sy, Georgi:1974yf, Georgi:1974my, Fritzsch:1974nn} (GUTs) are theoretical frameworks that seek to unify three fundamental forces of the Standard Model (SM) of elementary particle physics ---  the electromagnetic, weak, and strong interactions --- into a single one. Within these frameworks, the fundamental constituents of matter, namely quarks and leptons, are also partially or completely unified. One of the hallmark predictions of GUTs is thus that protons are not absolutely stable. Proton decay is, therefore, a crucial signature of GUTs that offers a potential avenue to test these theories, even though the relevant energy scales are far beyond the direct reach with current or any foreseeable experimental capabilities.

The discovery of proton decay would be a groundbreaking shift in the field of particle physics and an important development in our understanding of the universe. The search for proton decay has consequently motivated development of humongous detectors of unprecedented sensitivity, such as Super-Kamiokande and its successor, Hyper-Kamiokande (Hyper-K). It is worthwhile to point out that other forthcoming large-volume neutrino experiments, specifically JUNO and DUNE, are expected to probe proton decay lifetimes with an efficacy nearing that of the maximum achievable Hyper-K reach.

Hyper-K is the next-generation neutrino observatory and is scheduled to start observation in 2027. Remarkably, due to its enhanced sensitivity and a larger detection volume, Hyper-K will significantly improve the chances of observing proton decay with partial lifetimes as high as $10^{35}$\,years within a decade of operation. Owing to these advancements on the experimental side, it is timely to conduct an in-depth evaluation of the potential for observing proton decay in predictive and simple unified theories, which is the prime objective of the current manuscript.

The Georgi-Glashow model~\cite{Georgi:1974sy} played a pivotal role in the early development of GUTs. However, this model, consisting of only an adjoint Higgs to break the GUT symmetry and a fundamental Higgs to break the electroweak symmetry, is not compatible with experimental data due to its simplicity. More precisely, within this model (\textit{i}) gauge couplings do not unify, (\textit{ii}) one cannot accommodate neutrino masses,  and (\textit{iii}) there is a degeneracy between the down-type quark and charged lepton masses that contradicts the data. 

It is, therefore, essential to extend the Georgi-Glashow model to overcome the limitations listed above. While making such an extension, it is desirable to keep the model as simple as possible and, more importantly, predictive. In light of this, one of the simplest and realistic GUTs, based on $SU(5)$ gauge symmetry, was recently proposed in Ref.~\cite{Dorsner:2019vgf}. This minimal realization utilizes only the first five lowest-lying representations of dimensions 5, 10, 15, 24, and 35. It has a limited number of model parameters that makes it a highly predictive setup. Intriguingly, apart from two superheavy gauge bosons, only a single scalar leptoquark participates in proton decay mediation. This specific feature, coupled with the intricate interplay of model parameters necessary to replicate the experimentally observed masses and mixings of charged fermions and neutrinos, yields predictions regarding both the unification scale and the rates of various proton decay modes. The aim of this manuscript is to thoroughly examine the relationship between these decay modes and to arrive at robust predictions poised for potential observation in the forthcoming Hyper-K experiment.  

To concisely highlight our findings, we present the following summary. Our study infers that the decay of a proton into $p \to \pi^0 e^+$ indicates gauge boson mediation, which also opens a possibility of observing the $p \to \eta^0 e^+$ decay. Conversely, a proton decay that follows the $p \to K^+ \overline\nu$ route suggests an exchange of a scalar leptoquark, raising a potential to detect the $p \to \pi^0 \mu^+$ decay. The concurrent observation of $p \to \pi^0 e^+$ and $p \to K^+ \overline\nu$ decays might amplify the chances of seeing the $p \to \pi^0 \mu^+$ decay due to an interference effect. Our model forecasts that decay modes $p \to \pi^+ \overline\nu$, $p \to \eta^0 \mu^+$, $p \to K^0 e^+$, and $p \to K^0 \mu^+$ will remain out of reach in the coming decades, regardless of the dominant decay mechanism.

Our manuscript is organized as follows. We present main features of the model, such as the particle content interactions and symmetry breaking intricacies, in Sec.~\ref{SEC:model}. Since the SM fermion mass generation relies on three different mechanisms for its viability, we discuss relevant subtleties of these mechanisms in Sec.~\ref{SEC:fermionmass}. The proton decay predictions, for both the gauge boson and scalar leptoquark mediations, are discussed at length in Sec.~\ref{SEC:protondecay}. Predictions of our model, with regard to gauge coupling unification and proton decay signatures for all eight two-body decay processes and both types of mediations, are presented in Sec.~\ref{SEC:predictions}. We briefly conclude in Sec.~\ref{SEC:gcu}.

%%%%%%%%%%%%%%%%%%%%%%%%%%%%%%%%%%%%%%%%%%%%%%%
%%%%%%%%%%%%%%%%%%%%%%%%%%%%%%%%%%%%%%%%%%%%%%%
\section{Model}
\label{SEC:model}
The model under consideration contains scalars ($5_H$, $24_H$, and $35_H$), fermions ($\overline 5_{Fi}$, $10_{Fi}$, and $15_F+\overline{15}_F$), and gauge fields ($24_G$), where $i=1,2,3$. The full scalar and fermionic content of this $SU(5)$ model, including its decomposition under the SM gauge group, is summarized in Table~\ref{table:fields}. 
%%%%%%%%%%%%%%%%%%%%%%%%%%%%%%%%%%%%
\begin{table}
\begin{center}
\begin{tabular}{| c | c | c | c | c | c |}
\hline
\multicolumn{3}{|c|}{SCALARS} 
& 
\multicolumn{3}{c|}{FERMIONS} 
\\ 
\hline
\hline
$SU(5)$ & Standard Model & $(b_3,b_2,b_1)$ & $SU(5)$ & Standard Model & $(b_3,b_2,b_1)$\\\hline
\hline
 & $\Lambda_1\left(1,2,+\frac{1}{2}\right)$ & $\left(0,\frac{1}{6},\frac{1}{10}\right)$ & & $L_i\left(1,2,-\frac{1}{2}\right)$ & $\left(0,1,\frac{3}{5}\right)$\\
\raisebox{2ex}[0pt]{$\Lambda=5_H$} & $\Lambda_3\left(3,1,-\frac{1}{3}\right)$ & $\left(\frac{1}{6},0,\frac{1}{15}\right)$ & \raisebox{2ex}[0pt]{$F_i={\overline{5}_F}_i$} & $d_i^c\left(\overline{3},1,+\frac{1}{3}\right)$ & $\left(1,0,\frac{2}{5}\right)$\\ 
\hline
 & $\phi_0 \left(1,1,0\right)$ & $\left(0,0,0\right)$ & & $Q_i\left(3,2,+\frac{1}{6}\right)$ & $\left(2,3,\frac{1}{5}\right)$\\
 & $\phi_1 \left(1,3,0\right)$ & $\left(0,\frac{1}{3},0\right)$ & $T_i={10_F}_i$ & $u_i^c\left(\overline{3},1,-\frac{2}{3}\right)$ & $\left(1,0,\frac{8}{5}\right)$\\
$\phi=24_H$ & $\phi_3\left(3,2,-\frac{5}{6}\right)$ & $\left(\frac{1}{6},\frac{1}{4},\frac{5}{12}\right)$ &  & $e_i^c \left(1,1,+1\right)$ & $\left(0,0,\frac{6}{5}\right)$\\\cline{4-6}
 & $\phi_{\overline{3}} \left(\overline{3},2,+\frac{5}{6}\right)$ & $\left(\frac{1}{6},\frac{1}{4},\frac{5}{12}\right)$ &  & $\Sigma_1(1,3,+1)$ & $\left(0,\frac{4}{3},\frac{6}{5}\right)$\\
  & $\phi_8 \left(8,1,0\right)$ & $\left(\frac{1}{2},0,0\right)$ & $\Sigma=15_F$ & $\Sigma_3\left(3,2,+\frac{1}{6}\right)$ & $\left(\frac{2}{3},1,\frac{1}{15}\right)$\\\cline{1-3}
  & $\Phi_1  \left(1,4,-\frac{3}{2}\right)$ & $\left(0,\frac{5}{3},\frac{9}{5}\right)$ & & $\Sigma_6\left(6,1,-\frac{2}{3}\right)$ & $\left(\frac{5}{3},0,\frac{16}{15}\right)$\\\cline{4-6}
  & $\Phi_3  \left(\overline{3},3,-\frac{2}{3}\right)$ & $\left(\frac{1}{2},2,\frac{4}{5}\right)$ & & $\overline{\Sigma}_1\left(1,3,-1\right)$ & $\left(0,\frac{4}{3},\frac{6}{5}\right)$\\
 \raisebox{2ex}[0pt]{$\Phi=35_H$} & $\Phi_6  \left(\overline{6},2,+\frac{1}{6}\right)$ & $\left(\frac{5}{3},1,\frac{1}{15}\right)$ & $\overline{\Sigma}=\overline{15}_F$ & $\overline{\Sigma}_3\left(\overline{3},2,-\frac{1}{6}\right)$ & $\left(\frac{2}{3},1,\frac{1}{15}\right)$\\
 & $\Phi_{10}  \left(\overline{10},1,+1\right)$ & $\left(\frac{5}{2},0,2\right)$ &  & $\overline{\Sigma}_6\left(\overline{6},1,+\frac{2}{3}\right)$ & $\left(\frac{5}{3},0,\frac{16}{15}\right)$\\
\hline  
\end{tabular}
\end{center}
\caption{Particle content of the model, its decomposition under the SM gauge group, and associated $\beta$-function coefficients. $i=1,2,3$ represents a generation index.}
\label{table:fields}
\end{table}
%%%%%%%%%%%%%%%%%%%%%%%%%%%%%%%%%%%%

The scalar potential of the model reads
\begin{align}
\mathcal{L}_V= &
-\mu^2_\Lambda \left(\Lambda^*_\alpha\Lambda^\alpha\right) +\lambda^\Lambda_0  \left( \Lambda^*_\alpha\Lambda^\alpha  \right)^2 + \mu_1 \Lambda^*_\alpha\Lambda^\beta \phi^\alpha_\beta +  \lambda^\Lambda_1  \left(\Lambda^*_\alpha\Lambda^\alpha\right) \left(\phi^\beta_\gamma\phi_\beta^\gamma\right) + \lambda^\Lambda_2 \Lambda^*_\alpha\Lambda^\beta \phi_\beta^\gamma\phi_\gamma^\alpha
\nonumber \\ &
-\mu^2_\phi  \left(\phi^\beta_\gamma\phi_\beta^\gamma\right)
+\mu_2 \phi^\alpha_\beta\phi^\beta_\gamma \phi^\gamma_\alpha
+\lambda^\phi_0 \left( \phi^\beta_\gamma\phi_\beta^\gamma \right)^2
+\lambda^\phi_1  \phi^\alpha_\beta\phi^\beta_\gamma \phi^\gamma_\delta\phi^\delta_\alpha 
+\mu^2_\Phi \left(\Phi^{*\alpha \beta\gamma}\Phi_{\alpha \beta\gamma}\right)
\nonumber \\ &
+\lambda^\Phi_0 \left( \Phi^{*\alpha \beta\gamma}\Phi_{\alpha \beta\gamma}\right)^2
+\lambda^\Phi_1 \Phi^{*\alpha \beta\gamma}\Phi_{\alpha \beta\delta} \Phi^{*\delta\rho\sigma}\Phi_{\rho\sigma\gamma}
+\lambda_0 \left(\Phi^{*\alpha \beta\gamma}\Phi_{\alpha \beta\gamma}\right) \left(\phi^\delta_\rho\phi_\delta^\rho\right)
\nonumber \\ &
+\lambda_0^\prime \left(\Phi^{*\alpha \beta\gamma}\Phi_{\alpha \beta\gamma}\right)  \left(\Lambda^*_\rho\Lambda^\rho\right)
+\lambda_0^{\prime \prime} \Phi^{*\alpha \beta\gamma}\Phi_{\beta\gamma\delta}\Lambda^\delta\Lambda^*_\alpha
+\mu_3 \Phi^{*\alpha \beta\gamma}\Phi_{\beta\gamma\delta}\phi^\delta_\alpha
\nonumber \\ & 
+\lambda_1 \Phi^{*\alpha \beta\gamma}\Phi_{\alpha \delta\rho} \phi^\delta_\beta\phi^\rho_\gamma
+\lambda_2 \Phi^{*\alpha \beta\rho}\Phi_{\alpha \beta\delta} \phi^\gamma_\rho\phi^\delta_\gamma 
+ \left\{\lambda^\prime \Lambda^\alpha\Lambda^\beta\Lambda^\gamma \Phi_{\alpha\beta\gamma} +\mathrm{h.c.}\right\}, \label{eq:potential}
\end{align}
where, again, $\Lambda=5_H$, $\phi=24_H$, $\Phi=35_H$, and $\alpha, \beta, \gamma, \delta, \sigma, \rho =1,2,3,4,5$ are used to denote associated $SU(5)$ indices. 

The complete Yukawa part of the Lagrangian reads
\begin{align}
\mathcal{L}_Y= &
\left\{Y^u_{ij}T^{\alpha\beta}_iT^{\gamma\delta}_j\Lambda^\rho \epsilon_{\alpha\beta\gamma\delta\rho} 
+Y^d_{ij}T^{\alpha\beta}_iF_{\alpha j} \Lambda^{\ast}_\beta 
+Y^{a}_{i}\Sigma^{\alpha\beta}F_{\alpha i}  \Lambda^{\ast}_\beta
+Y^{b}_{i} \overline{\Sigma}_{\beta\gamma}F_{\alpha i} \Phi^{*\alpha\beta\gamma} \right.
\nonumber \\ &
\left. +Y^{c}_{i} T^{\alpha\beta}_i  \overline{\Sigma}_{\beta\gamma}\phi^\gamma_\alpha+\mathrm{h.c.}\right\}
+M_{\Sigma} \overline{\Sigma}_{\alpha\beta} \Sigma^{\alpha\beta}
+y \overline{\Sigma}_{\alpha\beta} \Sigma^{\beta\gamma} \phi_\gamma^\alpha , \label{eq:yukawa}
\end{align}
where we specify contractions in both the $SU(5)$ and flavor spaces. $Y^u$ and $Y^d$ are, in general, arbitrary $3\times 3$ complex Yukawa coupling matrices, $Y^a$, $Y^b$, and $Y^c$ are complex Yukawa coupling vectors of length 3, whereas $y$ is just a real number. The relevant matrix elements are denoted with $Y^u_{ij}$, $Y^d_{ij}$, $Y^{a}_{i}$, $Y^{b}_{i}$, $Y^{c}_{i}$, and $y$, where, again, $i,j=1,2,3$ represent generation indices.

We note that it is possible, without loss of generality, to use unitary field rotations of representations $\overline{5}_{Fi}$ and ${10_F}_j$ to go into a specific basis where $Y^d$ is a diagonal matrix with real and positive entries. We will use this basis for subsequent discussion and our numerical analysis. Also, all physical interactions at the SM level that feature $Y^u$ are always proportional to linear combinations $Y^u_{ij} + Y^u_{ji}$ that are manifestly symmetric in the flavor space. One can thus either redefine $Y^u$ to be a symmetric matrix and keep track of factors of two associated with that redefinition or continue to treat $Y^u$ as an arbitrary $3\times 3$ complex matrix. We opt for the latter approach.

The $SU(5)$ symmetry of the model is broken directly down to the SM gauge group $SU(3) \times SU(2) \times U(1)$ when $\phi_0 \in 24_H$ acquires a specific vacuum expectation value (VEV). The SM is subsequently broken at the electroweak scale by a VEV of $\Lambda_1 \in 5_H$ down to $SU(3) \times U(1)_\mathrm{em}$. This two-step process can be schematically represented as
\begin{align}
SU(5) \xrightarrow[]{\langle 24_H\rangle} SU(3) \times SU(2) \times U(1) \xrightarrow[]{\langle 5_H\rangle} SU(3) \times U(1)_\mathrm{em}  \;.  \end{align}

The relevant VEVs, for our study, are
\begin{align}
&
\langle 24_H \rangle =v_{24} \mathrm{diag}\left( -1, -1, -1, 3/2, 3/2 \right), \label{eq:24vev}
\\& \langle 5_H \rangle = (0 \quad 0 \quad 0 \quad 0 \quad v_{5}/\sqrt{2})^T \label{eq:5vev},
\end{align}
where $v_5 \approx 246$\,GeV reproduces the masses of the SM gauge boson fields $W_\mu^\pm$ and $Z_\mu$. The masses of the superheavy gauge bosons $X_\mu^{\pm 4/3} \in 24_G$ and $Y_\mu^{\pm 1/3} \in 24_G$, with the VEV of $24_H$ given via Eq.~\eqref{eq:24vev}, turn out to be
\begin{align}
M_X=M_Y=\sqrt{\frac{25}{8}} g_\mathrm{GUT} v_{24},
\end{align}
where $g_\mathrm{GUT}$ is a gauge coupling constant at the unification scale. 

One especially convenient feature of $SU(5)$ is that the gauge coupling unification scale $M_\mathrm{GUT}$ can be identified, for all practical purposes, with the mass of proton decay mediating gauge bosons $X_\mu^{\pm 4/3}$ and $Y_\mu^{\pm 1/3}$. We can thus set $M_X=M_Y\equiv M_\mathrm{GUT}$ when we discuss proton decay signatures via gauge mediation. Also, our model has only one scalar leptoquark $\Lambda_3 \in 5_H$ of mass $M_{\Lambda_3}$ that mediates proton decay.

Note that our discussion of the symmetry breaking procedure does not include possible VEV of an electrically neutral component of $\phi_1 \in 24_H$ that, if present, contributes to masses of $W_\mu^\pm$ gauge boson fields~\cite{Dorsner:2007fy}. We also omit a VEV of an electrically neutral component of $\Phi_1 \in 35_H$ in our symmetry breaking discussion. These VEVs are expected to be much smaller than the electroweak scale and can thus be safely neglected from our symmetry breaking discussion.

If $SU(5)$ symmetry is exact, all the SM multiplets within a given $SU(5)$ representation are degenerate in mass. This, of course, is no longer true after we spontaneously break it. We accordingly find two particular relations between the masses of the SM multiplets residing in $35_H$ and $15_F+\overline{15}_F$ that are especially relevant for the gauge coupling unification study. Namely, an analysis of the potential in Eq.~\eqref{eq:potential}, after the first stage of symmetry breaking, stipulates that
\begin{equation}
M^2_{\Phi_{10}}=M^2_{\Phi_1}-3M^2_{\Phi_3}+3 M^2_{\Phi_6}, \label{eq:mass_relation_b}
\end{equation}
whereas the last two terms in Eq.~\eqref{eq:yukawa} yield
\begin{equation}
M_{\Sigma_6}=2 M_{\Sigma_3}-M_{\Sigma_1}. \label{eq:mass_relation_a}
\end{equation}

Our one-loop level gauge coupling unification analysis reveals that fields $\Phi_{10}$ and $\Phi_1$ need to be degenerate in mass and heavy, i.e., of the order of $10^{11}$\,GeV, whereas $\Phi_3$ and $\Phi_6$ also need to be degenerate in mass but light, i.e., of the order of $10$\,TeV or less, if $M_\mathrm{GUT}$ is to be at, or exceed, existing experimental limits on partial proton decay lifetimes. The mass spectrum of fields in $\Phi=35_H$ that yields the highest possible value of $M_\mathrm{GUT}$ is thus $M_{\Phi_3}=M_{\Phi_6} \ll M_{\Phi_{10}}=M_{\Phi_1}$.

Also, under the assumption that parameters $M_{\Sigma_6}$, $M_{\Sigma_3}$, and $M_{\Sigma_1}$ are all positive, our gauge coupling unification study shows that fields $\Sigma_6$, $\Sigma_3$, and $\Sigma_1$ want to be almost perfectly mass-degenerate. This happens for the following two reasons. Firstly, Eq.~\eqref{eq:mass_relation_a} shows that $M_{\Sigma_3}$ is an arithmetic mean of $M_{\Sigma_1}$ and $M_{\Sigma_6}$. Secondly, the only field that has the correct $\beta$-function coefficients to increase $M_\mathrm{GUT}$ is $\Sigma_3$, whereas $\Sigma_1$ and $\Sigma_6$ can only decrease it, if they reside below $M_\mathrm{GUT}$. So, there are only two possible non-degenerate scenarios in agreement with Eq.~\eqref{eq:mass_relation_a} that one needs to consider to fully understand this issue. One scenario is when $M_{\Sigma_1} \ll M_{\Sigma_3} \sim M_{\Sigma_6}$ and the other is when $M_{\Sigma_6} \ll M_{\Sigma_3} \sim M_{\Sigma_1}$. It is clear that in both instances the GUT scale would be smaller that in the case when $M_{\Sigma_6} \sim M_{\Sigma_3} \sim M_{\Sigma_1}$ since the lightest state, i.e., either $\Sigma_1$ or $\Sigma_6$, would reduce the maximal allowed value of the GUT scale, whereas the other two states cannot subsequently affect $M_\mathrm{GUT}$ since they would enter the gauge coupling running at practically the same threshold, at some higher scale, and would thus, together with the lightest state, form a full $SU(5)$ multiplet. A preferred scenario that yields the largest $M_\mathrm{GUT}$ value and generates unification of gauge coupling constants turns out to be a scenario when $M_{\Sigma_6}$ is at most a factor of 10 below $M_{\Sigma_3}$, while $M_{\Sigma_3}$ is a factor $1/2$ below $M_{\Sigma_1}$, all in accordance with Eq.~\eqref{eq:mass_relation_a}. 

The fact that $\Sigma_6$, $\Sigma_3$, and $\Sigma_1$ want to be almost perfectly mass-degenerate implies that their common mass scale, in that regime, is not constrained by the gauge coupling unification requirement, at all. It turns out, however, that phenomenological viability of neutrino masses requires $M_{\Sigma_1}$ to be of the order of $M_{\Phi_1}$, as we demonstrate later on. 

We implement one-loop level gauge coupling unification analysis in order to find the largest possible value of unification scale $M_\mathrm{GUT}$ and associated value of $\alpha_\mathrm{GUT}=g^2_\mathrm{GUT}/(4 \pi)$ for fixed values of $M_{\Phi_1}$ and $M_{\Sigma_1}$. Again, since $M_{\Phi_1} = M_{\Phi_{10}}$ and $M_{\Sigma_1} \sim M_{\Sigma_3} \sim M_{\Sigma_6}$, the only parameters we need to vary within our unification analysis are the masses of $\Phi_3, \Phi_6 \in 35_H$, $\phi_1, \phi_8 \in 24_H$, and $\Lambda_3 \in 5_H$, while the relevant SM input parameters are $M_Z=91.1876$\,GeV, $\alpha_S(M_Z)=0.1193\pm0.0016$, $\alpha^{-1}(M_Z)=127.906\pm0.019$, and
$\sin^2 \theta_W(M_Z)=0.23126\pm0.00005$~\cite{Agashe:2014kda}. Since we also require $M_{\Lambda_3}$ to be at, or above, $3 \times 10^{11}$\,GeV due to the scalar mediated proton decay requirements~\cite{Dorsner:2012uz}, the only masses one actually needs to vary in our approach are $M_{\Phi_3}$, $M_{\Phi_6}$, $M_{\phi_1}$, and $M_{\phi_8}$.

The procedure we perform is as follows. First we specify a lower limit on the masses of the new physics states to establish a connection between the most accessible scale of new physics and $M_\mathrm{GUT}$. This lower limit is simply denoted with $M=\min(M_J)$, where $J(=\Phi_1, \Phi_3, \Phi_6, \Phi_{10}, \Sigma_1, \Sigma_3, \Sigma_6, \phi_1, \phi_8, \Lambda_3$) is set at $1$\,TeV and $10$\,TeV. For example, the $M\geq 1$\,TeV scenario simply means that the new physics states $J$ cannot reside below $1$\,TeV scale. Again, this lower limit, in practice, only affects masses of $\Phi_3$, $\Phi_6$, $\phi_1$, and $\phi_8$ as all other fields need to be much more massive.

We, next, produce a discrete set of values for $M_{\Phi_1}$ and $M_{\Sigma_1}$, where we take $\log_{10} (M_{\Phi_1}/1\,\mathrm{GeV}) \in [10.2,12.9]$ and $\log_{10} (M_{\Sigma_1}/1\,\mathrm{GeV}) \in [8.4,14.2]$ with the lattice spacing of 0.1 in these $\log$ units in both directions within $M_{\Phi_1}$-$M_{\Sigma_1}$ plane. The particular choice of ranges for $M_{\Phi_1}$ and $M_{\Sigma_1}$ is dictated by a need to generate high enough $M_\mathrm{GUT}$ to avoid rapid proton decay and to simultaneously obtain viable neutrino mass scale. For each of these $(M_{\Phi_1},M_{\Sigma_1})$ points we freely vary remaining $M_J$ masses until we find maximal possible $M_\mathrm{GUT}$. Note that our procedure, at every point in $M_{\Phi_1}$-$M_{\Sigma_1}$ plane, generates $\alpha_\mathrm{GUT}$, $M_\mathrm{GUT}$, and associated mass spectrum. We also account for conditions given with Eqs.~\eqref{eq:mass_relation_b} and~\eqref{eq:mass_relation_a} when generating $M_\mathrm{GUT}$, where we discard all those points with $M_\mathrm{GUT} \leq 6 \times 10^{15}$\,GeV due to gauge mediated proton decay requirements to save on computing time in subsequent steps. We show in Fig.~\ref{fig:parspace} contours of constant value of $\alpha_\mathrm{GUT}$ (blue dot-dashed lines) and $M_\mathrm{GUT}/(10^{15}\,\mathrm{GeV})$ (red solid lines) that we obtain with this procedure for $M\geq 1$\,TeV and $M\geq 10$\,TeV scenarios. 

One thing to note is that the value of $M_\mathrm{GUT}$ is independent of $M_{\Sigma_1}$ due to almost perfect mass-degeneracy of fields in $15_F+\overline{15}_F$, in agreement with our previous discussion. Also, to produce $M_\mathrm{GUT}$ values, as given in Fig.~\ref{fig:parspace}, we need to have four multiplets to be exactly at $M$ scale. These multiplets are $\Phi_3, \Phi_6 \in 35_H$ and $\phi_1, \phi_8 \in 24_H$. If the mass of any of these four multiplets is simply shifted up, the unification scale $M_\mathrm{GUT}$ would go down. This means that we are discussing the most conservative scenario with regard to the proton decay signatures via gauge boson exchange. Also, one can treat scale $M$, for all practical purposes, as a geometric mean of masses $M_{\Phi_3}$, $M_{\Phi_6}$, $M_{\phi_1}$, and $ M_{\phi_8}$ when one studies impact of its change on $M_\mathrm{GUT}$.

The second step of the numerical analysis is to run the masses and mixing parameters of the SM charged fermions from $M_Z$ to $M_\mathrm{GUT}$ using the factual new physics mass spectrum associated with every unification point that was obtained in the previous step. The charged fermion mass renormalization group running is performed at the one-loop level~\cite{Arason:1991ic}. We separate the gauge coupling unification study from the running of the SM charged fermion parameters since the latter provides feedback to the former only at the two-loop level, whereas the former impacts the latter already at the one-loop level. A summary of experimentally measured observables with the associated $1\,\sigma$ uncertainties of both the charged and neutral fermion sectors at low-energy scale is provided in Table~\ref{tab:input}.  
\begin{table}[th!]
\centering
\footnotesize
\resizebox{0.9\textwidth}{!}{
\begin{tabular}{|c|c|c|c|}
\hline
\pbox{5cm}{\vspace{5pt}$m(M_{Z})$ (GeV)\vspace{5pt}} & Fit input  &  $\theta^{\rm{CKM},\rm{PMNS}}_{ij}$ \& $\delta^{\rm{CKM}}$ \& $\Delta m^2_{ij}$ (eV$^2$) & Fit input  \\
\hline
\hline
$m_{u}/10^{-3}$ & $1.158 \pm 0.392$ &$\sin\theta^{\rm{CKM}}_{12}$ & $0.2254\pm 0.00072$ \\ \hline
$m_{c}$ & $0.627 \pm 0.019$ &$\sin\theta^{\rm{CKM}}_{23}/10^{-2}$ & $4.207\pm 0.064$  \\ \hline
$m_{t}$ & $171.675 \pm 1.506$ &$\sin\theta^{\rm{CKM}}_{13}/10^{-3}$ & $3.640\pm 0.130$  \\ \hline
$m_{d}/10^{-3}$ & $2.864 \pm 0.286$ &$\delta^{\rm{CKM}}$ & $1.208\pm0.054$ \\ \hline
$m_{s}/10^{-3}$ & $54.407 \pm 2.873$ &$\Delta m^{2}_{21}/10^{-5}$ &7.425$\pm$0.205  \\ \hline
$m_{b}$ & $2.854 \pm 0.026$ & $\Delta m^{2}_{3\ell}/10^{-3}$ &2.515$\pm$0.028 \\ \hline
$m_{e}/10^{-3}$ & $0.486576$ &$\sin^{2}\theta^{\rm{PMNS}}_{12}/10^{-1}$  &3.045$\pm$0.125   \\ \hline
$m_{\mu}$ & $0.102719$ &$\sin^{2}\theta^{\rm{PMNS}}_{23}$ &0.554$\pm$0.021  \\ \hline
$m_{\tau}$ & $1.74618$  &$\sin^{2}\theta^{\rm{PMNS}}_{13}/10^{-2}$ &$2.224\pm 0.065$  \\ \hline
\end{tabular}
}
\caption{Experimental measurements related to charged fermions~\cite{Antusch:2013jca} and neutrinos for normal ordering~\cite{Esteban:2020cvm} with 1\,$\sigma$ uncertainties.}
\label{tab:input}
\end{table}
Due to a minimal impact of the running of neutrino observables from $M_Z$ to $M_\mathrm{GUT}$, we opt to always use corresponding low-energy scale values in our numerical analysis.

%%%%%%%%%%%%%%%%%%%%%%%%%%%%%%%%%%%%%%%%%%%%%%%
%%%%%%%%%%%%%%%%%%%%%%%%%%%%%%%%%%%%%%%%%%%%%%%
\section{Fermion Masses}
\label{SEC:fermionmass}

The SM fermion mass generation, in our model, relies on three different mechanisms for its viability. We discuss relevant subtleties of these mechanisms in what follows.

\subsection{Charged fermion sector}
%%%%%%%%%%%%%%%%%%%%%%%%%%%%%%%%%%%%%%%%%%%%%%%
We first revisit the charged fermion mass generation without inclusion of the vector-like fermion corrections for clarity of exposition. Recall, $i$-th generation of the SM fermions, in the Georgi-Glashow $SU(5)$, is entirely embedded in $\overline{5}_{Fi}$ and $10_{Fi}$~\cite{Georgi:1974sy}. For example, the first generation is 
\begin{align}
\overline{5}_{F1}=\begin{pmatrix}
d^C_1\\d^C_2\\d^C_3\\e\\ -\nu_e
\end{pmatrix},\;\;\;10_{F1}=\frac{1}{\sqrt{2}} \begin{pmatrix}
0&u^C_3&-u^C_2&u_1&d_1\\
-u^C_3&0&u^C_1&u_2&d_2\\
u^C_2&-u^C_1&0&u_3&d_3\\
-u_1&-u_2&-u_3&0&e^C\\
-d_1&-d_2&-d_3&-e^C&0
\end{pmatrix},
\end{align}
where all the fields are left-handed. 

Once the electroweak symmetry is broken via VEV of Eq.~\eqref{eq:5vev}, the first two terms in Eq.~\eqref{eq:yukawa} generate the following mass matrices for the up-type quarks, charged leptons, and down-type quarks:
\begin{align}
 M_U &=\sqrt{2}v_5\left(Y^u+Y^{u\,T}\right),\label{eq:uncorrected_U}\\
M_E &= \frac{v_5}{2}Y^{d\,T},\label{eq:uncorrected_E}\\
M_D &= \frac{v_5}{2}Y^d. \label{eq:uncorrected_D}
\end{align}
We again note that we work in a basis where $Y^d$ is a real diagonal matrix. Clearly, Eqs.~\eqref{eq:uncorrected_E} and~\eqref{eq:uncorrected_D} predict that $m_e=m_d$, $m_\mu=m_s$, and $m_\tau=m_b$, where these mass relations hold at the unification scale $M_\mathrm{GUT}$. 

To generate experimentally observed mismatch between the masses of the down-type quarks and charged leptons we clearly need to introduce corrections to either $M_E$ or $M_D$, or both. Required corrections, in our model, stem from interactions of the SM charged fermions in $\overline{5}_{Fi}$ and ${10_F}_j$ with the vector-like fermions in $15_F + \overline{15}_F$~\cite{Shafi:1999rm,Shafi:1999ft,Oshimo:2009ia,Dorsner:2019vgf,Dorsner:2021qwg,Antusch:2023jok,Antusch:2023mqe}. Namely, once both the $SU(5)$ and electroweak symmetries are broken, fermion submultiplets with the same transformation properties under the $SU(3) \times U(1)_\mathrm{em}$ gauge group that reside in $\overline{5}_{Fi}$, ${10_F}_j$, $15_F$, and $\overline{15}_F$ will mix. The mixing terms of interest arise from the third, fourth, and fifth contractions of Eq.~\eqref{eq:yukawa} and read
\begin{align}
\mathcal{L}_Y\supset &
-Y^{a}_{i}\;\left( \Sigma^0 \nu_i  \frac{h^0}{\sqrt{2}}
+
\Sigma^{e^C} e^-_i \frac{v_5}{\sqrt{2}}
+
\Sigma^d d^C_i \frac{v_5}{\sqrt{2}}
\right)
\nonumber\\&
-Y^{b}_{i}\; \overline \Sigma^0  \nu_i \frac{\Phi^0_{\mathrm{Re}}}{\sqrt{2}}
-\frac{5v_{24}}{4} Y^{c}_{i}\; \left(
d_i \overline\Sigma^d + u_i \overline\Sigma^u
\right),  \label{eq:corrections}
\end{align}
where we introduce the electric charge eigenstate fields $\Sigma^0, \Sigma^{e^C} \in \Sigma_1(1,3,1)$ and $\Sigma^u, \Sigma^d\in \Sigma_3(3,2,1/6)$. Moreover, $\Phi^0_{\mathrm{Re}}$ denotes the real part of $\Phi_1 \in 35_H$, whereas $h^0 \in \Lambda_1 \in 5_H$ denotes what will primarily be the SM Higgs boson. We stress that $\Phi^0_{\mathrm{Re}}$ and $h^0$, strictly speaking, are not exact mass eigenstates as these two states mix, as we discuss later. The masses of associated linear combinations of $\Phi^0_{\mathrm{Re}}$ and $h^0$ will subsequently be denoted with $M_{\Phi_1}$ and $M_h$, where $M_{\Phi_1} \gg M_h$. Also, since $M_{\Sigma_1}$ needs to be of the order of $M_{\Phi_1}$ that is heavy and $M_{\Sigma_1} \approx M_{\Sigma_3} \approx M_{\Sigma_6}$, we can safely neglect any mass splitting between fermions $\Sigma^0$ and $\Sigma^{e^C}$ or $\Sigma^u$ and $\Sigma^d$ that is induced through the mixing with the SM fermions.

Finally, Eqs.~\eqref{eq:uncorrected_U},~\eqref{eq:uncorrected_E},~\eqref{eq:uncorrected_D} and~\eqref{eq:corrections}, when put together, yield $4\times 4$ mass matrices for the charged fermions that read
\begin{align}
\mathcal{L}_Y \supset &
\begin{pmatrix}u_i&\Sigma^u\end{pmatrix}
\begin{pmatrix}
\begin{array}{c|c}
\sqrt{2} v_5 (Y^{u}_{ij}+Y^{u}_{ji}) & -\frac{5v_{24}}{4} Y^c_i \\ \hline
0 &M_{\Sigma_3}
\end{array}
\end{pmatrix}
\begin{pmatrix}u^C_j\\\overline{\Sigma}^u\end{pmatrix} 
  \nonumber\\& 
+\begin{pmatrix}d_i&\Sigma^d\end{pmatrix}
\begin{pmatrix}
\begin{array}{c|c}
\frac{1}{2}v_5 Y^d_{ij} & -\frac{5v_{24}}{4} Y^c_i\\ \hline
\frac{1}{\sqrt{2}}v_5 Y^a_j & M_{\Sigma_3}
\end{array}
\end{pmatrix}
\begin{pmatrix}d^C_j\\\overline{\Sigma}^d\end{pmatrix}
\nonumber\\&
+\begin{pmatrix}e_i&\overline{\Sigma}^{e^C}\end{pmatrix}
\begin{pmatrix}
\begin{array}{c|c}
\frac{1}{2}v_5 Y^d_{ji} & \frac{1}{\sqrt{2}}v_5 Y^a_i\\ \hline 0 &M_{\Sigma_1}
\end{array}
\end{pmatrix}
\begin{pmatrix}e^C_j\\\Sigma^{e^C}\end{pmatrix}\,.   \label{MASS-FORMULA}
\end{align}

Again, we are interested in a scenario where $M_{\Sigma_i}\gg v_5$. In this particular limit, the $3\times 3$ mass matrices of the charged fermions of the SM take the form
\begin{align}
&M_U=\left( \mathbb{I}+\epsilon^2\;Y^c{Y^c}^{\dagger} \right)^{-\frac{1}{2}} \;\sqrt{2} v_5(Y^{u}+Y^{u\,T}),  \label{massu}
\\
&M_E=\frac{1}{2} v_5 Y^{d\,T} \label{masse},
\\
&M_D=\left( \mathbb{I}+\epsilon^2\;Y^c{Y^c}^{\dagger} \right)^{-\frac{1}{2}}   \left( M_E^T +\frac{1}{\sqrt{2}} v_5\; \epsilon\; Y^cY^a  \right),  \label{massd}
\end{align}
where we define
\begin{align}
\epsilon= \frac{5}{4} \frac{v_{24}}{M_{\Sigma_3}}.   
\end{align}

Our numerical analysis reveals that $\mathbb{I}+\epsilon^2\;Y^c{Y^c}^{\dagger} \approx \mathbb{I}$ to a great accuracy. This guarantees that $M_U=M_U^T$ and makes transparent that a mismatch between the down-type quark and charged lepton masses originates solely from a single rank-one matrix with elements proportional to the product $Y^c_i Y^a_j$. Also, a nice feature of this set-up is that the values of charged fermion masses directly determine the entries of diagonal matrix $Y^d$ via Eq.~\eqref{masse}.

\subsection{Neutrino sector}
%%%%%%%%%%%%%%%%%%%%%%%%%%%%%%%%%%%%%%%%%%%%%%%
Neutrino masses receive contributions at both the tree- and one-loop levels~\cite{Babu:2009aq,Bambhaniya:2013yca}. However, in the viable parameter space of our model, the tree-level contribution to the neutrino masses can be completely neglected when compared to the latter one. We, nevertheless, quote it for completeness. It reads
\begin{equation}
\label{eq:tree}
(M_N)^\mathrm{tree}_{ij}= - \frac{\lambda^\prime v^4_5}{ 4 M_{\Sigma_1} M^2_{\Phi_1}} \left(Y^a_iY^b_j+Y^b_iY^a_j\right). 
\end{equation}
Eq.~\eqref{eq:tree} implies that the need to have rather heavy $\Phi_1$, i.e., $M_{\Phi_1} \sim 10^{11}$\,GeV, in order to unify gauge coupling constants of the SM at high enough scale $M_\mathrm{GUT}$, does not allow for large enough neutrino mass scale even for electroweak scale $M_{\Sigma_1}$.
 
\begin{figure}[t!]
\centering
\includegraphics[width=0.47\textwidth]{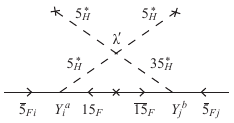}
\includegraphics[width=0.47\textwidth]{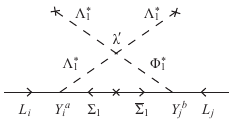}
\caption{The Feynman diagrams of the leading order contribution towards Majorana neutrino masses at the $SU(5)$ (left panel) and the SM (right panel) levels.}
\label{fig:diagram}
\end{figure}

The leading contribution to neutrino masses is of the one-loop level topology and is shown in Fig.~\ref{fig:diagram}. (For other radiative neutrino mass models within the $SU(5)$ setup, see, for example, Refs.~\cite{Wolfenstein:1980sf,Barbieri:1981yw,Perez:2016qbo,Kumericki:2017sfc,Saad:2019vjo,Klein:2019jgb}.)
%%%%%%%%%%%%%%%%%%
%%%%%%%%%%%%%%%%%%
A necessary ingredient to generate non-zero masses for the neutrinos, as can be seen from Fig.~\ref{fig:diagram}, is the very last term in Eq.~\eqref{eq:potential} that leads to a mixing between $h^0 \in 5_H$ and a field $\Phi^0_{\mathrm{Re}} \in 35_H$. This term reads 
\begin{align}
\mathcal{L}_V\supset     \lambda^\prime\; \Lambda^\alpha\Lambda^\beta\Lambda^\gamma \Phi_{\alpha\beta\gamma} +\mathrm{h.c.} \supset 2\times \lambda^\prime\;\frac{v_5^2}{4} h^0  \Phi^0_{\mathrm{Re}},
\end{align}
where we explicitly show the $SU(5)$ origin of the relevant SM interaction. 

From the above term, the mixing angle $\theta$ between $h^0$ and $\Phi^0_{\mathrm{Re}}$ is given by
\begin{align}
\sin 2\theta=  \frac{\lambda^\prime v^2_5}{M^2_{\Phi_1}-M^2_h}\;,
\end{align}
where, again, $M_{\Phi_1}$ and $M_h$ denote physical masses of linear combinations of $h^0 \in \Lambda_1 \in 5_H$ and $\Phi^0_{\mathrm{Re}} \in \Phi_1 \in 35_H$. Here, $M_h$ is the mass of the SM Higgs boson. If we properly take into account all the mixing and loop factors, we derive the following neutrino mass formula:  
\begin{align}
(M_N)^\mathrm{loop}_{ij}&= \frac{\lambda^\prime v_5^2}{64 \pi^2} (Y^a_iY^b_j+Y^b_iY^a_j) 
\frac{M_{\Sigma_1}}{M^2_{\Phi_1}-M^2_h} \bigg\{
\frac{M^2_{\Phi_1}\ln \frac{M^2_{\Sigma_1}}{M^2_{\Phi_1}}}{M^2_{\Sigma_1}-M^2_{\Phi_1}}
-
\frac{M^2_h\ln \frac{M^2_{\Sigma_1}}{M^2_h}}{M^2_{\Sigma_1}-M^2_h}
\bigg\}.
\end{align}

In the limit when $M_{\Phi_1},M_{\Sigma_1} \gg M_h$, we find
\begin{equation}
\label{eq:m_neutrino}
(M_N)^\mathrm{loop}_{ij}= m_0 (Y^a_iY^b_j+Y^b_iY^a_j),
\end{equation}
where we introduce dimensionful parameter $m_0$ that sets the neutrino mass scale. It reads
\begin{equation}
\label{eq:m0}
m_0= \frac{\lambda^\prime v_5^2}{64\pi^2} \frac{M_{\Sigma_1}}{M^2_{\Sigma_1}-M^2_{\Phi_1}} \ln \frac{M^2_{\Sigma_1}}{M^2_{\Phi_1}}.
\end{equation}
The mass dependence of Eq.~\eqref{eq:m0} clearly demonstrates that it is possible to have satisfactorily large value of $m_0$ for $M_{\Phi_1} \sim 10^{11}$\,GeV as long as $M_{\Sigma_1} \gg M_h$.

We note that both the tree- and one-loop level contributions towards neutrino mass matrix are proportional to $Y^a_iY^b_j+Y^b_iY^a_j$ combination. This insures that the lightest of the neutrinos is a massless particle, which is consistent with the current neutrino oscillation data. We can thus write
\begin{equation}
(M_N)_{ij}
= m_0 \left(
Y^a_i Y^b_j+Y^b_i Y^a_j
\right)= (N\;
 \mathrm{diag}(0,m_2,m_3)\;
N^T)_{ij}\,, 
\label{eq:nu}
\end{equation}
where $N$ is a unitary matrix and $m_2$ and $m_3$ are neutrino mass eigenstates. $N$ can actually be written as
\begin{equation}
N=\mathrm{diag} (e^{i\gamma_1}, e^{i\gamma_2}, e^{i\gamma_3}) V_\mathrm{PMNS}^*,
\label{Nmatrix}
\end{equation}
where $V_\mathrm{PMNS}$ is the Pontecorvo-Maki-Nakagawa-Sakata (PMNS) unitary mixing matrix with three mixing angles, two Majorana phases, and one CP violating Dirac phase. 

We can use results of Refs.~\cite{Cordero-Carrion:2018xre,Cordero-Carrion:2019qtu} to obtain entries of $Y^a$ and $Y^b$ from Eq.~\eqref{eq:nu} as a function of $N$, $m_2$, and $m_3$. Namely, we find, for the normal neutrino mass ordering scenario, the following expressions for ${Y^a}^T$ and ${Y^b}^T$:
\begin{equation}
\label{eq:perturbativity}
Y^{a\,T}=\frac{1}{\rho\sqrt{2}} \begin{pmatrix}
i\;r_2\;N_{12}+r_3\;N_{13}\\
i\;r_2\;N_{22}+r_3\;N_{23}\\
i\;r_2\;N_{32}+r_3\;N_{33}
\end{pmatrix},\;\;
Y^{b\,T}=\frac{\rho}{\sqrt{2}} \begin{pmatrix}
-i\;r_2\;N_{12}+r_3\;N_{13}\\
-i\;r_2\;N_{22}+r_3\;N_{23}\\
-i\;r_2\;N_{32}+r_3\;N_{33}
\end{pmatrix},
\end{equation}
where $r_2=\sqrt{m_2/m_0}$, $r_3=\sqrt{m_3/m_0}$, and $\rho$ is an unknown parameter that accounts for the fact that the matrix elements of $Y^a$ and $Y^b$ are always featured as products $Y^a_i Y^b_j$ in Eq.~\eqref{eq:nu}. Since $V_\mathrm{PMNS}$ has three phases, i.e., one CP violating Dirac phase and two Majorana phases, unitary matrix $N$ has, all in all, six phases. Hence in Eq.~\eqref{eq:perturbativity} we have six arbitrary phases. This is consistent with the process of trading six real parameters and six phases from $Y^a$ and $Y^b$ for three PMNS angles, three neutrino masses, and six phases during the inversion procedure.
 
We can use experimental input from neutrino sector to directly constrain viable parameter space of the model. Namely, since the neutrino mass scale parameter $m_0$ depends solely on $M_{\Phi_1}$, $M_{\Sigma_1}$, and $\lambda^\prime$, we can establish a part of parameter space in $M_{\Phi_1}$-$M_{\Sigma_1}$ plane where $m_0$ is simply too small to accommodate experimentally observed neutrino mass-square differences with perturbative couplings. Namely, if we demand that the dimensionless couplings do not exceed value of one, we can exclude part of parameter space with requirement that $m_0\leq m_3/2$. This naive approach excludes parameter space to the right of a green curve labeled with $1$ in both panels of Fig.~\ref{fig:parspace}, where we set $|\lambda^\prime|=1$. To be precise, green curves in both panels of Fig.~\ref{fig:parspace} are contours of constant $m_0$ in units of $\sqrt{\Delta m^2_{31}}/2$. For example, green curve labeled $10$ can be interpreted in two different ways. If, for example, we set $|\lambda^\prime|=1$, the Yukawa coupling product $\max (|Y^a_i Y^b_j|)$, along that curve, needs to be $10^{-1}$. Or, if we demand that the Yukawa couplings do not exceed value of one, the value of $|\lambda^\prime|$ is $10^{-1}$ along green curve labeled with 10. 

We can do a more accurate analysis with regard to perturbativity of the Yukawa couplings in the neutrino sector. Namely, since $Y^a$ and $Y^b$ entries, as given via Eq.~\eqref{eq:perturbativity}, are completely determined via six unknown phases and $\lambda^\prime$, we can vary these phases and demand that both $\max(|Y^a_i|) \leq 1$ and $\max(|Y^b_i|) \leq 1$, where $i=1,2,3$, to insure perturbativity for fixed values of $M_{\Phi_1}$, $M_{\Sigma_1}$, and $|\lambda^\prime|=1$. If we do that, we find that the parameter space to the right of the outermost dashed black curve in both panels of Fig.~\ref{fig:parspace} is ruled out for any choice of aforementioned six phases. In other words, it is impossible to have perturbative Yukawa couplings in the neutrino sector for any point in $M_{\Phi_1}$-$M_{\Sigma_1}$ plane to the right of outermost dashed black curve. Similarly, for any choice of these six phases we always find perturbative solution to the left of the innermost dashed black curve. Clearly, green curve labeled with $1$ is a good approximation of a more accurate numerical study. The region between two dashed black curves represents a region where one can have satisfactory numerical fit of neutrino masses with perturbative Yukawa couplings but only for very specific choices of six phases that enter $Y^a$ and $Y^b$ via Eq.~\eqref{eq:perturbativity}.

%%%%%%%%%%%%%%%%%%%%%%%%%%%%%%%%%%%%%%%%%%%%%%%
\subsection{Fermion mass fit}
\label{mass_fit}
The $3\times 3$ mass matrices $M_U$, $M_E$, $M_D$, and $M_N$ of the SM fermions, as given in Eqs.~\eqref{massu},~\eqref{masse},~\eqref{massd}, and~\eqref{eq:nu}, respectively, can be diagonalized via 
\begin{align}
&M_U= U_L M_U^\mathrm{diag} U_R^\dagger\,,  \label{eq:massu_a}
\\
&M_D= D_L M_D^\mathrm{diag} D_R^\dagger\,,  \label{eq:massd_a}
\\
&M_E = E_L M_E^\mathrm{diag} E_R^\dagger\,, \label{eq:masse_a}
\\
&M_N=N M_N^\mathrm{diag} N^T\,,
\label{eq:massnu_a}
\end{align}
where $U_L$, $U_R$, $D_L$, $D_R$, $E_L$, $E_R$, and $N$ are unitary matrices that implement transition from flavor to mass-eigenstate basis.

Our numerical study yields $D_L$, $D_R$, and $N$, as we discuss in detail later on, whereas the features of the model stipulate that
\begin{align}
\label{eq:test_2}
U_L&= D_L \mathrm{diag} (1, e^{i\eta_1}, e^{i\eta_2}) V_{\mathrm{CKM}}^T\,\mathrm{diag} (e^{i\kappa_1}, e^{i\kappa_2}, e^{i\kappa_3}) \equiv D_L D(\eta)V_{\mathrm{CKM}}^T D(\kappa)\\
    U_R&= U_L^*\,\mathrm{diag} (e^{i\xi_1}, e^{i\xi_2}, e^{i\xi_3}) \equiv U_L^* D(\xi),
\label{eq:test_3}
\\
\label{eq:test_4}
E_L&=\mathbb{I}\,,\\
\label{eq:test_5}
E_R&=\mathbb{I}\,,
\end{align}
where $V_\mathrm{CKM}$ is the Cabibbo-Kobayashi-Maskawa (CKM) mixing matrix, which contains one CP violating Dirac phase $\delta^\mathrm{CKM}$. We introduce in Eqs.~\eqref{eq:test_2} and~\eqref{eq:test_3} convenient notation for diagonal phase matrices $D(\eta)$, $D(\kappa)$, and $D(\xi)$. Note, also, that it is a symmetric nature of $M_U$ matrix that relates $U_L$ and $U_R$, as given in Eq.~\eqref{eq:test_3}.

To performe numerical fit, we use charged lepton masses at $M_\mathrm{GUT}$ as an input to determine the diagonal matrix $Y^d$ via $Y^d= 2\;\mathrm{diag}(m_e,m_\mu,m_\tau)/v_5$. Since the down-type quark mass matrix of Eq.~\eqref{massd} and the neutrino mass matrix of Eq.~\eqref{eq:nu} share a common Yukawa coupling row matrix $Y^a$, a combined fit of these two sectors is necessary. Our numerical fit accordingly minimizes a $\chi^2$ function 
\begin{align}
\chi^2= \sum_{j=1}^8\left(\frac{T_j-O_j}{E_j}\right)^2,  \end{align}
where $T_j$, $O_j$, and $E_j$ represent theoretical prediction, measured central value, and experimental $1\,\sigma$ error for the observable $j$, respectively. The index $j$ runs over the down-type quark masses and all five measured observables in the neutrino sector.  

We determine the Yukawa coupling matrices $Y^{a}$, $Y^{b}$, and $Y^{c}$ by fitting them against three down-type quark masses, two neutrino mass-squared differences, and three mixing angles in the neutrino sector. It is important to note that the CP-violating Dirac phase and the two Majorana phases in the neutrino sector have not yet been experimentally measured. These, together with the entries of $D_L$ and $D_R$, are actually an output of our fit. 

We conduct a comprehensive scan over all viable unification points presented in Fig.~\ref{fig:parspace}, imposing the condition of perturbativity for the relevant couplings, i.e., $\max(|Y^a_i|), \max(|Y^b_i|), \max(|Y^c_i|), |\lambda^\prime | \leq 1$. Additionally, we employ the criterion $\chi^2/n \leq 1$ to deem a fit as good, where $n(=8)$ represents the number of fitted observables. It is noteworthy that not all unification points that allow for a numerically good fit to the fermion sector successfully pass the proton decay test.

Our comprehensive numerical fit encompassing both the down-type quark and neutrino sectors highlights a limitation in accommodating the inverted neutrino mass ordering within this model. Specifically, the hierarchical and diagonal nature of the down-type quark mass matrix is dictated by the charged lepton Yukawa coupling matrix $Y^d$. Since the matrix elements $(M_D)_{ij}$ depend on a linear combination of $(Y^d)_{ij}$ and $Y^c_i Y^a_j$, it becomes evident that both $Y^a$ and $Y^c$ must manifestly be hierarchical row matrices to yield a satisfactory fit to the data. However, achieving this hierarchical row structure is untenable when considering the inverted ordering of neutrino masses. In the inverted scenario, the entries in the first row and the first column of the neutrino mass matrix $M_N$ typically possess comparable magnitudes, while the lower $2\times 2$ block is required to be somewhat smaller. Working in the mass eigenstate basis for the charged leptons, this necessity imposes a constraint that all entries in $Y^a$ must be of the same order, contradicting the hierarchical arrangement needed in the down-type quark sector. This tension arises directly from the simplicity of our model, leading to the prediction that neutrinos must exhibit the normal mass ordering.

The fitting procedure, applied under the assumption of the normal ordering of neutrino masses, enables numerical determination of three unitary rotation matrices $D_L$, $D_R$, and $N$, along with the Yukawa couplings for the charged leptons and the down-type quarks. To comprehensively calculate partial lifetimes for various proton decay modes arising from both gauge boson and scalar mediations, it is essential to have knowledge of the unitary matrices $U_L$ and $U_R$ that diagonalize the up-type quark mass matrix and its associated Yukawa couplings. A nice feature of our approach is the ability to express both $U_L$ and $U_R$ in terms of $D_L$ and $V_\mathrm{CKM}$, as specified in Eqs.~\eqref{eq:test_2} and~\eqref{eq:test_3}, along with eight additional phases.

In summary, the model accurately accommodates charged lepton masses, up-type quark masses, and CKM parameters. Additionally, we conduct a combined numerical fit for the neutrino mass parameters, down-type quark masses, and PMNS parameters, recognizing their intrinsic interdependence. A pivotal outcome of the fit for the proton decay considerations lies in the unitary transformations $U_L$, $U_R$, $D_L$, and $D_R$, with the first two matrices featuring five and three unknown phases, respectively, and numerical determination of all Yukawa couplings of the model. There are, all in all, eight additional phases that our fit cannot determine, i.e., $\eta_1$, $\eta_2$, $\kappa_1$, $\kappa_2$, $\kappa_3$, $\xi_1$, $\xi_2$, and $\xi_3$, where, as we show next, only $\eta_1$ and $\eta_2$ are relevant for the study of proton decay signatures. This puts us in a perfect position to produce an in-depth anatomy of proton decay signatures of this model.

%%%%%%%%%%%%%%%%%%%%%%%%%%%%%%%%%%%%%%%%%%%%%%%
%%%%%%%%%%%%%%%%%%%%%%%%%%%%%%%%%%%%%%%%%%%%%%%

\section{Proton Decay}
\label{SEC:protondecay}
The main idea behind our proton decay analysis is to accurately identify the most dominant channels for both types of mediators that are present in our model and compare associated predictions for partial lifetimes with current experimental limits and future expectations based on a ten-year period of data taking at Hyper-K. A summary of the best experimental limits for all eight two-body proton decay channels as well as expectations for future sensitivities at Hyper-K, if and when available, is accordingly given in Table~\ref{tab:present_future}. (JUNO Collaboration expects to reach a limit of $9.6\times 10^{33}$\,years for $p\rightarrow K^+\overline{\nu}$ after a ten-year period of data taking~\cite{JUNO:2022qgr}.) 
\begin{table} 
\centering
\begin{tabular}{|c|c|c|}
  \hline
decay channel & $\tau_p$ current bound & $\tau_p$ future sensitivity \\   
\hline \hline 
$p\rightarrow\pi^0e^+$ & $2.4\times 10^{34}$\,years~\cite{Super-Kamiokande:2020wjk} & $7.8\times 10^{34}$\,years~\cite{Hyper-Kamiokande:2018ofw} \\ 
$p\rightarrow\pi^0\mu^+$  & $1.6\times 10^{34}$\,years~\cite{Super-Kamiokande:2020wjk} & $7.7\times 10^{34}$\,years~\cite{Hyper-Kamiokande:2018ofw}\\
$p\rightarrow \pi^+\overline{\nu}$  & $3.9\times 10^{32}$\,years~\cite{Super-Kamiokande:2013rwg} & - \\ 
$p\rightarrow\eta^0e^+$ & $1.0\times 10^{34}$\,years~\cite{Super-Kamiokande:2017gev} & $4.3\times 10^{34}$\,years~\cite{Hyper-Kamiokande:2018ofw} \\ 
$p\rightarrow\eta^0\mu^+$  & $4.7\times 10^{33}$\,years~\cite{Super-Kamiokande:2017gev} & $4.9\times 10^{34}$\,years~\cite{Hyper-Kamiokande:2018ofw} \\  
$p\rightarrow K^0e^+$  & $1.1\times 10^{33}$\,years~\cite{Brock:2012ogj} & - \\  
$p\rightarrow K^0\mu^+$  & $3.6\times 10^{33}$\,years~\cite{Super-Kamiokande:2022egr} & - \\ 
$p\rightarrow K^+\overline{\nu}$  & $8.0\times 10^{33}$\,years~\cite{Okumura:2020xfs} &  $3.2\times10^{34}$\,years~\cite{Hyper-Kamiokande:2018ofw}\\ 
  \hline
\end{tabular}
\caption{Current limits on partial proton decay lifetimes for all two-body decay processes and future expectations for a ten-year period of data taking at 90\,\%\,C.L..}
\label{tab:present_future}
\end{table}
We furthermore provide, in what follows, analytic expressions for partial decay widths to address phase dependence of our results. We do that separately for the gauge boson and scalar leptoquark proton decay mediations. Recall, relevant gauge bosons are $X_\mu^{+ 4/3},Y_\mu^{+ 1/3} \in (\overline{3},2,+5/6) \in 24_G$ with a common mass $M_\mathrm{GUT}$, whereas scalar leptoquark is $\Lambda_3 (3,1,-1/3) \in 5_H$ with mass $M_{\Lambda_3}$.

%%%%%%%%%%%%%%%%%%%%%%%%%%%%%%%%%%%%%%%%%%%%%%%
\subsection{Proton decay via gauge bosons}
To evaluate relevant decay widths we resort to a formalism presented in Refs.~\cite{FileviezPerez:2004hn,Nath:2006ut}. The $d=6$ level operators of interest are
\begin{align}
    O (e^C_{\alpha}, d_{\beta}) &= k^2 c(e^C_{\alpha}, d_{\beta})\,\epsilon_{ijk}\,\overline{u^C_{i}}\,\gamma^{\mu}\,u_{j}\,\overline{e^{C}_{\alpha}}\,\gamma_{\mu}\,d_{k\beta},\label{eq:G1}\\
    O (e_{\alpha}, d^C_{\beta}) &= k^2 c(e_{\alpha}, d^C_{\beta})\,\epsilon_{ijk}\,\overline{u^C_{i}}\,\gamma^{\mu}\,u_{j}\,\overline{d^{C}_{k \beta}}\,\gamma_{\mu}\,e_{\alpha},\label{eq:G2}\\
    O (\nu_l, d_{\alpha}, d^C_{\beta}) &= k^2 c(\nu_l, d_{\alpha}, d^C_{\beta})\,\epsilon_{ijk}\,\overline{u^C_{i}}\,\gamma^{\mu}\,d_{j \alpha}\,\overline{d^{C}_{k \beta}}\,\gamma_{\mu}\,\nu_l,\label{eq:G3}
\end{align}
where $k^2=g^2_\mathrm{GUT}/M^2_\mathrm{GUT}$, $ (e_1,e_2) \equiv (e,\mu)$, $ (d_1,d_2) \equiv (d,s)$, and $l=1,2,3$. 

We are primarily interested in derivation of explicit expressions for dimensionless coefficients $c(e^C_{\alpha}, d_{\beta})$, $c(e_{\alpha}, d^C_{\beta})$, and $c(\nu_l, d_{\alpha}, d^C_{\beta})$. The flavor dependent coefficients relevant for proton decays into charged leptons are
\begin{align}
    c(e_\beta , d^C)&= e^{-i\xi_1} (D^\dagger_R)_{1\beta},\\
    c(e_\beta^C , d)&= e^{-i\xi_1} \left[(D^*_L)_{\beta 1}+(V_{\mathrm{CKM}})_{11} (D_L^* D(\eta)^* V^\dagger_{\mathrm{CKM}})_{\beta 1} \right],\\
    c(e_{\beta}, s^C) &= e^{-i\xi_1} (D^\dagger_R)_{2\beta}\\
    c(e^C_{\beta}, s) &= e^{-i\xi_1} \left[ (D^*_L)_{\beta 2}+e^{i\eta_1} (V_{\mathrm{CKM}})_{12} (D^*_L D(\eta)^* V^\dagger_{\mathrm{CKM}})_{\beta 1} \right].
\end{align}
Clearly, a common multiplicative phase factor $e^{-i\xi_1}$ will disappear once one squares the amplitude in order to evaluate physical decay widths of interest. Also, the only parameters that are not provided by our numerical study of the gauge coupling unification and fermion mass fit are phases $\eta_1$ and $\eta_2$ in $D(\eta) \equiv \mathrm{diag} (1, e^{i\eta_1}, e^{i\eta_2})$.

To evaluate decay widths associated with $p\rightarrow \pi^+\overline{\nu}$ and $p\rightarrow K^+\overline{\nu}$ one needs to take into account the fact that the neutrino flavor is not an observable in proton decay experiments and thus needs to be summed over. We accordingly obtain  
\begin{align}
    \sum_{l=1}^3 |\,c(\nu_l , d, d^C)\,|^2 &= \left|\left(D(\xi)^*D(\kappa) V_{\mathrm{CKM}} D(\eta)\right)_{11}\right|^2= \left|(V_{\mathrm{CKM}})_{11}\right|^2,\label{eq:sum_1}\\
    \sum_{l=1}^3 \left| c(\nu_l , d, s^C)\right|^2 &= \left|\left(D(\xi)^*D(\kappa) V_{\mathrm{CKM}} D(\eta)\right)_{11}\right|^2 = \left|(V_{\mathrm{CKM}})_{11}\right|^2,\label{eq:sum_2}\\
    \sum_{l=1}^3 \left| c(\nu_l , s, d^C)\right|^2 &= \left|(V_{\mathrm{CKM}})_{12}\right|^2.\label{eq:sum_3}
\end{align}
The outcome of summations in Eqs.~\eqref{eq:sum_1},~\eqref{eq:sum_2}, and~\eqref{eq:sum_3} stipulates that the flavor dependence of $p\rightarrow \pi^+\overline{\nu}$ and $p\rightarrow K^+\overline{\nu}$ processes is completely determined by the CKM entries~\cite{FileviezPerez:2004hn}. 

Our analytical expressions show that the only two phases that are featured in gauge boson mediated proton decay into mesons and charged leptons are $\eta_1$ and $\eta_2$. In fact, even the dependence on $\eta_2$ can be neglected for all practical purposes, as we show later on. We stress that these analytical results will hold in any $SU(5)$ model where charged fermions are already in the mass eigenstate basis while the up-type quark mass matrix is symmetric in flavor space.  

Once the coefficients $c(e^C_{\alpha}, d_{\beta})$, $c(e_{\alpha}, d^C_{\beta})$, and $c(\nu_l, d_{\alpha}, d^C_{\beta})$ are determined, one can write down two-body proton decay widths and generate predictions of the model. For example, the most dominat decay width reads
\begin{align}
\Gamma (p \to \pi^0 e^+)&=  \frac{( m_p^2-m_\pi^2)^2}{m_p^3} \frac{\pi}{2} A_L^2 \frac{\alpha_\mathrm{GUT}^2}{M_\mathrm{GUT}^4} \nonumber\\
&\times \left(  A^2_{SL}\left| c(e^C,d) \langle\pi^0|(ud)_L u_L |p\rangle\right|^2  + A^2_{SR} \left| c(e,d^C) \langle
\pi^0|
(ud)_R u_L|p \rangle \right|^2 \right)\,, \label{decaywidthpi0e}
\end{align}
where $\matrixel{\pi^0}{(ud)_Lu_L}{p}=0.134\,\mathrm{GeV}^2$~\cite{Aoki:2017puj} and $\matrixel{\pi^0}{(ud)_Ru_L}{p}=-0.131\,\mathrm{GeV}^2$~\cite{Aoki:2017puj}. These matrix elements are evaluated with an error of the order of 10\,\% that we do not take into account in our numerical analysis. $A_L=1.2$ represents a long-distance coefficient~\cite{Nihei:1994tx}, whereas $A_{SL}$ and $A_{SR}$ are short-distance coefficients. $A_{SL}$ and $A_{SR}$ capture
the running of the proton decay operators from $M_\mathrm{GUT}$ down to $M_Z$ and are evaluated via~\cite{Buras:1977yy,Ellis:1979hy,Wilczek:1979hc}
\begin{equation*}
A_{SL(R)}=\prod_{i=1,2,3} \prod_I^{M_Z \leq M_I \leq M_\mathrm{GUT}}
\left[\frac{\alpha_i(M_{I+1})}{\alpha_i(M_I)}\right]^{\frac{\gamma_{L(R)i}}{\sum_J^{M_Z
\leq M_J \leq M_I} b^J_{i}}},\,
\gamma_{L(R)i}=(23(11)/20,9/4,2),
\end{equation*}
where indices $I$ and $J$ run through all the new physics states that reside below the unification scale. We evaluate $A_{SL}$ and $A_{SR}$ for every point in $M_\mathrm{\Phi_1}$-$M_\mathrm{\Sigma_1}$ plane for both $M \geq 1$\,TeV and $M \geq 10$\,TeV scenarios. All eight proton decay widths, due to the gauge boson exchange, are explicitly written down in Appendix~\ref{A}.

%%%%%%%%%%%%%%%%%%%%%%%%%%%%%%%%%%%%%%%%%%%%%%%
\subsection{Proton decay via scalar leptoquark}
The relevant $d=6$ operators for proton decay through scalar leptoquark $\Lambda_3$ are~\cite{Nath:2006ut}
\begin{align}
    O_H (d_\alpha, e_\beta) &= M^{-2}_{\Lambda_3} a(d_\alpha, e_\beta) \,u^T L \,C^{-1} \,d_\alpha\, u^T L\, C^{-1} \,e_\beta,\\
    O_H (d_\alpha, e^C_\beta) &= M^{-2}_{\Lambda_3} a(d_\alpha, e^C_\beta) \,u^T L \,C^{-1} \,d_\alpha\, e^{C^\dagger}_\beta L\, C^{-1} \,u^{C^*},\\
    O_H (d^C_\alpha, e_\beta) &= M^{-2}_{\Lambda_3} a(d^C_\alpha, e_\beta) \,d^{C^\dagger}_\alpha L \,C^{-1} \,u^{C^*}\, u^T L\, C^{-1} \,e_\beta,\\
    O_H (d^C_\alpha, e^C_\beta) &= M^{-2}_{\Lambda_3} a(d^C_\alpha, e^C_\beta) \,d^{C^\dagger}_\alpha L \,C^{-1} \,u^{C^*}\,e_\beta^{C^\dagger}\,  L\, C^{-1} \,u^{C^*},\\
    O_H (d_\alpha, d_\beta, \nu_l) &= M^{-2}_{\Lambda_3} a(d_\alpha, d_\beta, \nu_l)\, u^T\, L \, C^{-1}\, d_\alpha\,d^T_\beta \, L \, C^{-1}\, \nu_l,\\
    O_H (d_\alpha, d^C_\beta, \nu_l) &= M^{-2}_{\Lambda_3} a(d_\alpha, d^C_\beta, \nu_l)\,d_\beta^{C^{\dagger}}\, L\, C^{-1}\, u^{C^*}\, d^T_\alpha\,L\,C^{-1}\,\nu_l,
\end{align}
\noindent where C is a charge conjugation operator and $L=(1-\gamma_5)/2$. Again, it is dimensionless coefficients $a(d_\alpha, e_\beta)$, $a(d_\alpha, e^C_\beta)$, $a(d^C_\alpha, e_\beta)$, $a(d^C_\alpha, e^C_\beta)$, $a(d_\alpha, d_\beta, \nu_l)$, and $a(d_\alpha, d^C_\beta, \nu_l)$ that encapsulate flavor dependence of proton decay signatures we are interested in.

We find the following expressions~\cite{Dorsner:2012nq}
\begin{align}
    a (d_\alpha, e_\beta) &= e^{-i\xi_1} \frac{m^U_1 m^E_\beta}{v^2} \left(V_{\mathrm{CKM}} D(\eta) \right)_{1 \alpha} \left(V^*_{\mathrm{CKM}} D(\eta)^* D^\dagger_L \right)_{1\beta},\\
    a(d_\alpha, e^C_\beta) &= \frac{m_1^U}{m_\beta^E} a(d_\alpha, e_\beta),\\
    a(d_\alpha^C, e_\beta) &= e^{-i\xi_1}  \frac{m^E_\beta}{v^2} \left(D^\dagger_R M^\mathrm{diag}_E D_L D(\eta) V^T_{\mathrm{CKM}}\right)_{\alpha1} \left(V^*_{\mathrm{CKM}} D(\eta)^* D^\dagger_L\right)_{1\beta},\\
    a (d_\alpha^C, e^C_\beta) &= \frac{m_1^U}{m_\beta^E} a(d^C_\alpha, e_\beta),\\
    a (d_\alpha, d_\beta, \nu_l) &= e^{-i\xi_1+i\kappa_1} \frac{m_1^U}{v^2} \left(V_{\mathrm{CKM}} D(\eta) \right)_{1\alpha} \left(D^\dagger_L M_E^\mathrm{diag} N^* \right)_{\beta l},\\
    a (d_\alpha, d^C_\beta, \nu_l) & = \frac{e^{-i\xi_1+i\kappa_1}}{v^2} \left(D^\dagger_R M^\mathrm{diag}_E D_L D(\eta) V^T_{\mathrm{CKM}}\right)_{\beta1}\left(D^\dagger_L M^\mathrm{diag}_E N^* \right)_{\alpha l},
\end{align}
where $v=v_5/\sqrt{2}$, $m^U_1\equiv m_u$, and  $(m^E_1,m^E_2) \equiv (m_e,m_\mu)$. Clearly, the overall phase factors $e^{-i\xi_1}$ and $e^{-i\xi_1+i\kappa_1}$ will not be featured in any of the decay widths. The only phases that will leave an imprint on the partial decay lifetimes due to the scalar leptoquark mediation are, once again, $\eta_1$ and $\eta_2$.

To obtain decay widths for $p\rightarrow \pi^+\overline{\nu}$ and $p\rightarrow K^+\overline{\nu}$ processes due to scalar leptoquark exchange we need to sum over neutrino flavors. We accordingly find these contractions
\begin{align}   
\sum_{l=1}^3 a (d_\alpha, d_\beta, \nu_l) a^* (d_\gamma, d_\delta, \nu_l) &= \frac{m_1^{U^2}}{v^4} \left(V_{\mathrm{CKM}} D(\eta) \right)_{1\alpha}\nonumber\\
   &  \times \left(V^*_{\mathrm{CKM}} D(\eta)^* \right)_{1\gamma}\left(D^\dagger_L M^{\mathrm{diag}\,2}_E D_L \right)_{\beta\delta},\\
   \sum_{l=1}^3 a (d_\alpha, d^C_\beta, \nu_l) a^* (d_\gamma, d^C_\delta, \nu_l) &= \frac{1}{v^4} \left(D^\dagger_R M^\mathrm{diag}_E D_L D(\eta) V^T_{\mathrm{CKM}}\right)_{\beta1}\nonumber\\ & \times \left(D^T_R M^\mathrm{diag}_E D_L^* D(\eta)^* V^\dagger_{\mathrm{CKM}}\right)_{\delta 1}\left(D^\dagger_L M^{\mathrm{diag}\,2}_E D_L \right)_{\alpha\gamma},\\
   \sum_{l=1}^3 a (d_\alpha, d_\beta, \nu_l) a^* (d_\gamma, d^C_\delta, \nu_l) &= \frac{m_1^U}{v^4} \left(D^\dagger_L M^{\mathrm{diag}\,2}_E D_L \right)_{\beta\gamma}\nonumber\\ & \times \left(V_{\mathrm{CKM}} D(\eta) \right)_{1\alpha}\left(D^T_R M^\mathrm{diag}_E D_L^* D(\eta)^* V^\dagger_{\mathrm{CKM}}\right)_{\delta 1}.
\end{align}
It is clear that the parameters associated with the neutrino sector, including the PMNS matrix, are not featured in proton decay signatures via scalar leptoquark exchange.

Once all these coefficients are analytically determined we can write down two-body proton decay widths due to the scalar leptoquark mediation and generate predictions of our model. For example, the most constraining decay width is 
\begin{align}
    \Gamma (p\rightarrow K^+ \bar{\nu})
    & =\frac{(m_p^2-m^2_{K})^2}{32\pi m^3_p M_{\Lambda_3}^4} A_L^2 \nonumber \\
    &\times\sum_{i=1}^3\left|A_{SL} \left(a(s, d,\nu_i) \langle K^+ |(us)_L d_L|p \rangle + a(d, s,\nu_i) \langle K^+ |(ud)_L s_L|p\rangle\right) \right. \nonumber\\
    &\left. + A_{SR} \left(a(d, s^C,\nu_i) \langle K^+ |(us)_R d_L|p \rangle + a(s, d^C,\nu_i) \langle K^+ |(ud)_R s_L|p\rangle\right) \right|^2,
\end{align}
where the values of matrix elements $\langle K^+| (us)_{L,R} d_L |p \rangle$ and $\langle K^+| (ud)_{L,R} s_L |p \rangle$ are given in Table~\ref{tab:matrix_elements}. As before, $A_L$ represents a long-distance coefficient, whereas $A_{SL}$ and $A_{SR}$ are short-distance coefficients. The expressions of proton decay widths for all eight two-body decay modes can be found in Appendix~\ref{B}.

%%%%%%%%%%%%%%%%%%%%%%%%%%%%%%%%%%%%%%%%%%%%%%%
%%%%%%%%%%%%%%%%%%%%%%%%%%%%%%%%%%%%%%%%%%%%%%%
\section{Predictions}
\label{SEC:predictions}
We are finally in a position to place additional limits on otherwise viable parameter space of the model using our results for the gauge boson and scalar leptoquark proton decay mediations. 

\begin{figure}
\centering
\includegraphics[width=0.65\textwidth]{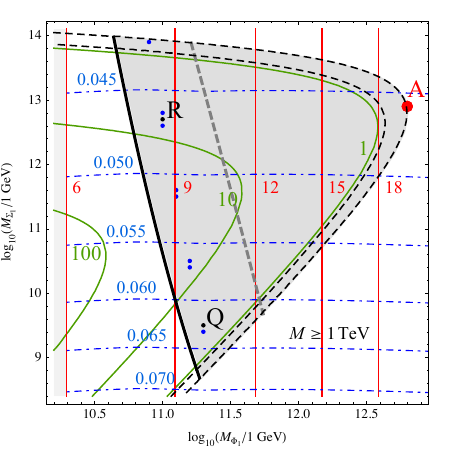}
\includegraphics[width=0.65\textwidth]{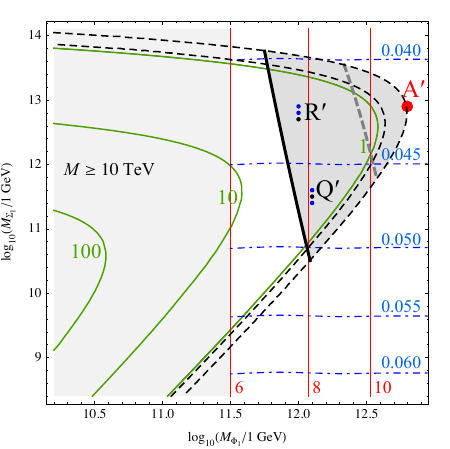}
\caption{\label{fig:parspace} Viable parameter space of the model (shaded in gray) for the $M \geq 1$\,TeV and $M\geq$ 10\,TeV scenarios with contours of constant values for $M_\mathrm{GUT}/(10^{15}\,\mathrm{GeV})$ (red solid lines), $\alpha_\mathrm{GUT}$ (blue dot-dashed lines), and $m_0/(\sqrt{\Delta m^2_{31}}/2)$ (green solid lines). For additional details, see the text.}
\end{figure}

The most relevant limit originates from the model prediction for the gauge mediatated $p \to \pi^0 e^+$ process given in Eq.~\eqref{decaywidthpi0e}. We present thus inferred proton decay limit in both panels of Fig.~\ref{fig:parspace} using a solid black line. Note that to the left of that line it is not possible to have partial proton decay lifetime of $p \to \pi^0 e^+$ that is in agreement with the current experimental limit for any choice of phases $\eta_1$ and $\eta_2$. In short, the shaded region between the proton decay bound (solid black line) and the outermost neutrino mass scale bound (dashed black curve) is viable with regard to all experimental input. We also show in Fig.~\ref{fig:parspace}, using gray dashed line, what one expects to be excluded with $p \to \pi^0 e^+$ non-observation after a ten-year period of data taking at Hyper-K~\cite{Hyper-Kamiokande:2018ofw}. 

Again, Fig.~\ref{fig:parspace} features contours of constant values for $M_\mathrm{GUT}/(10^{15}\,\mathrm{GeV})$ (red solid lines), $\alpha_\mathrm{GUT}$ (blue dot-dashed lines), and $m_0/(\sqrt{\Delta m^2_{31}}/2)$ (green solid lines) in $M_{\Phi_1}$-$M_{\Sigma_1}$ plane for $M \geq 1$\,TeV and $M \geq 10$\,TeV scenarios, where $\log_{10} (M_{\Phi_1}/1\,\mathrm{GeV}) \in [10.2, 12.9]$ and $\log_{10} (M_{\Sigma_1}/1\,\mathrm{GeV}) \in [8.4, 14.2]$. The lattice spacing in $M_{\Phi_1}$-$M_{\Sigma_1}$ plane for our analysis, in both directions, is 0.1 in units of $\log_{10} (M_{\Phi_1}/1\,\mathrm{GeV})$ and $\log_{10} (M_{\Sigma_1}/1\,\mathrm{GeV})$, respectively. Again, at each lattice point in $M_{\Phi_1}$-$M_{\Sigma_1}$ plane we evaluate mass spectrum that leads to maximal possible value of $M_\mathrm{GUT}$ and determine associated $\alpha_\mathrm{GUT}$, $A_{SL}$, and $A_{SR}$. The mass spectrum that generates $\max(M_\mathrm{GUT})$ is always such that $M_{\Phi_3} = M_{\Phi_6} = M_{\phi_1} = M_{\phi_8} = M$. We furthermore perform running of the SM charged fermion parameters at each lattice point and use associated values to perform numerical fit of all fermion masses. The outcome of the fit are all unitary transformations that provide transition from flavor to mass-eigenstate basis of the SM fermions and associated Yukawa coupling constants. Parameters that our procedure cannot determine are phases $\eta_1$, $\eta_2$, $\kappa_1$, $\kappa_2$, $\kappa_3$, $\xi_1$, $\xi_2$, and $\xi_3$, where  only $\eta_1$ and $\eta_2$ are relevant for the study of proton decay signatures.

In order to accomplish comparative study between gauge boson and scalar leptoquark mediation signatures for the partial proton decay lifetimes, we choose one particular point in Fig.~\ref{fig:parspace} for $M\geq$ 1\,TeV scenario to be our starting point. This point, with coordinates $(M_{\Phi_1},M_{\Sigma_1})=(10^{11.3}\,\mathrm{GeV},10^{9.5}\,\mathrm{GeV})$, is denoted Q and is chosen to be near the current proton decay bound rendered with solid black line in the upper panel of Fig.~\ref{fig:parspace}. We subsequently evaluate $\alpha_{\mathrm{GUT}}^2/M^4_{\mathrm{GUT}}$ at this point and find all other points in Fig.~\ref{fig:parspace} that satisfy criteria that the associated value of $\alpha_{\mathrm{GUT}}^2/M^4_{\mathrm{GUT}}$ at those points is within $\pm$2\,\% with respect to the value extracted for point Q. This procedure yields ten points of interest for $M\geq$ 1\,TeV scenario that are presented in the upper panel of Fig.~\ref{fig:parspace}.   

Once these ten points in $M_{\Phi_1}$-$M_{\Sigma_1}$ plane are known, we choose among them one point with large value of $\alpha_{\mathrm{GUT}}$ (Q$(10^{11.3}\,\mathrm{GeV},10^{9.5}\,\mathrm{GeV})$) and another one with small value of $\alpha_{\mathrm{GUT}}$ (R$(10^{11.0}\,\mathrm{GeV},10^{12.7}\,\mathrm{GeV})$), where R is singled out because it has value of $\alpha_{\mathrm{GUT}}^2/M^4_{\mathrm{GUT}}$ that is numerically almost identical to $\alpha_{\mathrm{GUT}}^2/M^4_{\mathrm{GUT}}$ evaluated at point Q. We subsequently evaluate at both of these points gauge boson mediated proton decay widths for all eight channels, where we also vary $\eta_1$ and $\eta_2$ phases to find maximal and minimal values for these decay widths. We furthermore evaluate scalar leptoquark mediated proton decay widths for all eight channels at both Q and R, where, once again, we vary $\eta_1$ and $\eta_2$ to capture phase dependence. To find decay widths associated with scalar leptoquark mediation, we fix the mass of $\Lambda_3$ to have a partial lifetime for $p \to K^+ \overline{\nu}$ as close as possible to the current experimental limit. We take $M_{\Lambda_3}=2.7 \times10^{12}$\,GeV for $M\geq$ 1\,TeV scenario. Again, this choice ensures that the scalar mediated proton decay signatures will be observed with certainty at future proton decay experiments. 
\begin{figure}
\centering
\includegraphics[width=0.49\textwidth]{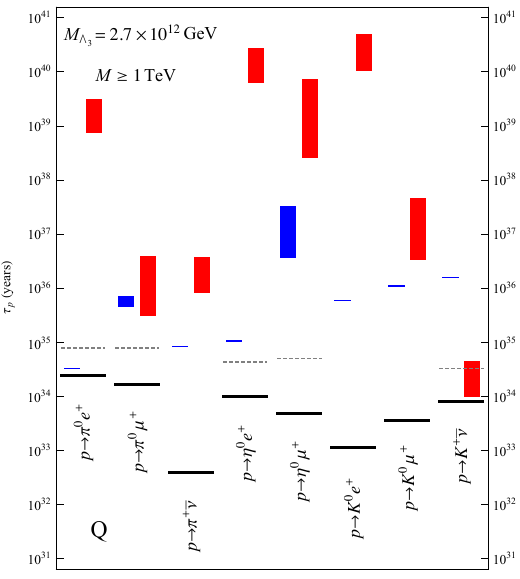}
\includegraphics[width=0.49\textwidth]{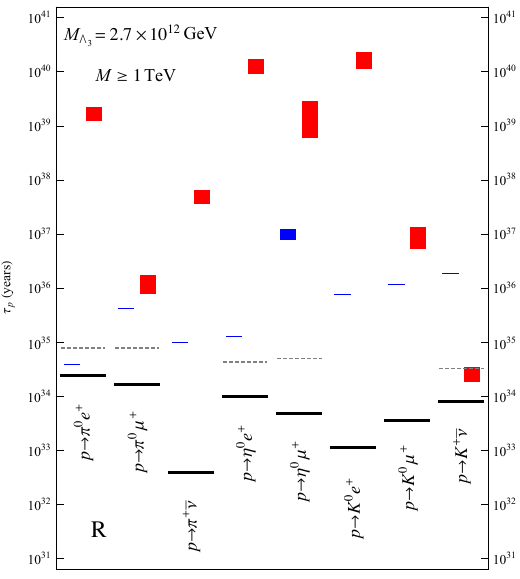}
\includegraphics[width=0.49\textwidth]{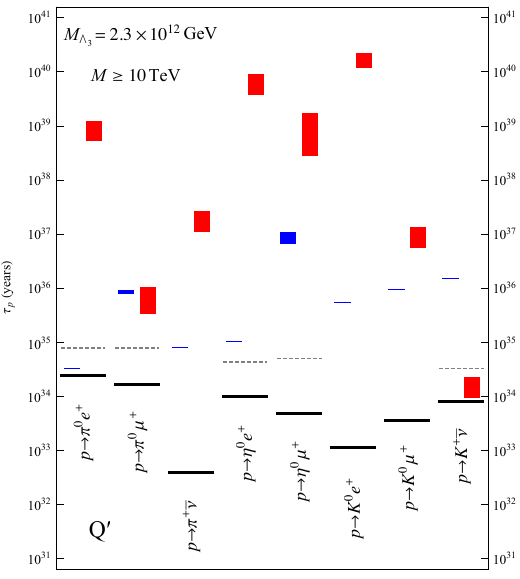}
\includegraphics[width=0.49\textwidth]{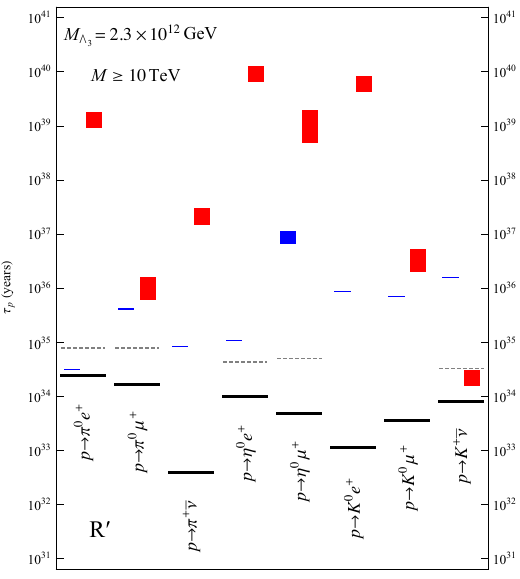}
\caption{\label{fig:GSsignatures} Proton decay signatures via gauge boson and scalar leptoquark mediations within $M\geq$ 1\,TeV and $M\geq$ 10\,TeV scenarios for four specific points and eight channels, as indicated. Black lines are current experimental limits, blue vertical bars are predictions for gauge boson mediation signatures, red vertical bars are predictions for the scalar leptoquark mediations, and gray dashed lines represent future experimental sensitivities after a ten-year period of data taking at 90\,\%\,C.L..}
\end{figure}

We preform the same procedure of selecting two particular points for $M \geq 10$\,TeV scenario of Fig.~\ref{fig:parspace}, where we label these points Q$^\prime(10^{12.1}\,\mathrm{GeV},10^{11.5}\,\mathrm{GeV})$ and R$^\prime(10^{12.0}\,\mathrm{GeV},10^{12.7}\,\mathrm{GeV})$. To evaluate scalar mediated proton decay signatures for points Q$^\prime$ and R$^\prime$ we set $M_{\Lambda_3}=2.3 \times10^{12}$\,GeV. Again, this choice of $M_{\Lambda_3}$ places prediction at point Q$^\prime$ for $p \to K^+ \overline{\nu}$ within an imminent reach of Hyper-K.   

We show signatures for both types of mediation side by side in Fig.~\ref{fig:GSsignatures} for all four points of interest, where blue bars are used for the gauge boson mediation partial lifetime predictions, red bars are used for predictions associated with scalar leptoquark mediation, current  experimental limits are represented by thin black lines, and gray dashed lines stand for future expectations after a ten-year period of data taking at 90\,\% C.L., if and where available. 

Upper panels in Fig.~\ref{fig:GSsignatures} are for proton decay signatures for points Q and R, whereas lower panels correspond to predictions for points Q$^\prime$ and R$^\prime$. Note that the uncertainties in proton decay widths associated with scalar leptoquark mediation are much larger than those associated with gauge boson mediation. Again, in both cases these uncertainties originate solely from variation of two phases $\eta_1$ and $\eta_2$. We quantify these flavor uncertainties in Table~\ref{tab:per_cent}, where we evaluate $\Delta = (\max (\tau_p)-\min (\tau_p))/\min (\tau_p)$ in \% at points Q, R, Q$^\prime$, and R$^\prime$ for each process and for both types of mediation.
\begin{table}[h]
\centering
\begin{tabular}{|c|c|c|c|c|c|c|c|c|}
\hline
$\Delta$ (\%) & \multicolumn{2}{c|}{gauge} 
& 
\multicolumn{2}{c|}{scalar} & \multicolumn{2}{c|}{gauge} 
& 
\multicolumn{2}{c|}{scalar}\\
\hline
\hline
decay channel &  $\Delta$(Q) & $\Delta$(R) & $\Delta$(Q) & $\Delta$(R) & $\Delta$(Q$^\prime)$ & $\Delta$(R$^\prime)$ & $\Delta$(Q$^\prime)$ & $\Delta$(R$^\prime)$\\
\hline 
\hline
$p \rightarrow \pi^0 e^+$ & 2 & 1 & 339 & 85 & 1 & 1 & 142 & 96\\
$p \rightarrow \pi^0 \mu^+$ & 54 & 6 & 1147 & 132 & 19 & 9 & 214 & 166\\
$p \rightarrow \pi^+ \overline{\nu}$ & 0 & 0 & 363 & 87 & 0 & 0 & 145 & 99\\
$p \rightarrow \eta^0 e^+$ & 3 & 1 & 358 & 90 & 1 & 1 & 146 & 97\\
$p \rightarrow \eta^0 \mu^+$ & 810 & 59 & 2743 & 382 & 70 & 81 & 513 & 313\\
$p \rightarrow K^0 e^+$ & 2 & 1 & 386 & 103 & 1 & 1 & 90 & 102\\
$p \rightarrow K^0 \mu^+$ & 5 & 1 & 1304 & 152 & 1 & 1 & 146 & 171\\
$p \rightarrow K^+ \overline{\nu}$ & 0 & 0 & 361 & 87 & 0 & 0 & 145 & 99\\
\hline 
\end{tabular}
\caption{Phase uncertainty $\Delta = (\max (\tau_p)-\min (\tau_p))/\min (\tau_p)$ in \% at points Q, R, Q$^\prime$, and R$^\prime$ for all eight channels and for both types of proton decay mediation.}
\label{tab:per_cent} 
\end{table}

One can see from Fig.~\ref{fig:GSsignatures} and Table~\ref{tab:per_cent} that gauge mediated proton decays $p \to \pi^+ \overline{\nu}$ and $p \to K^+\overline{\nu}$ do not exhibit any dependence on $\eta_1$ and $\eta_2$. Also, the prediction uncertainty is usually smaller for points with smaller value of $\alpha_\mathrm{GUT}$ for both types of mediation. One can furthermore observe that even though points Q and R have practically the same value of $\alpha_\mathrm{GUT}^2/M_\mathrm{GUT}^4$, the prediction for the $p \to \pi^0 e^+$ lifetime for point Q is closer to the experimental limit than for point R. This is due to the fact that the short-distance coefficients $A_{SL}$ and $A_{SR}$ are larger in the region where $\alpha_\mathrm{GUT}$ is larger. This effect can also be observed in the scalar leptoquark mediation signatures. For example, we find that $A_{SL}=3.41$ and $A_{SR}=3.14$ at point Q, whereas $A_{SL}=3.08$ and $A_{SR}=2.86$ at point R.

\begin{figure}[th!]
\centering
\includegraphics[width=0.49\textwidth]{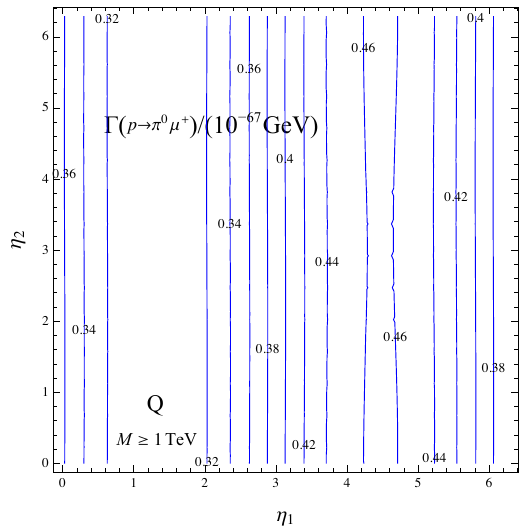}
\includegraphics[width=0.49\textwidth]{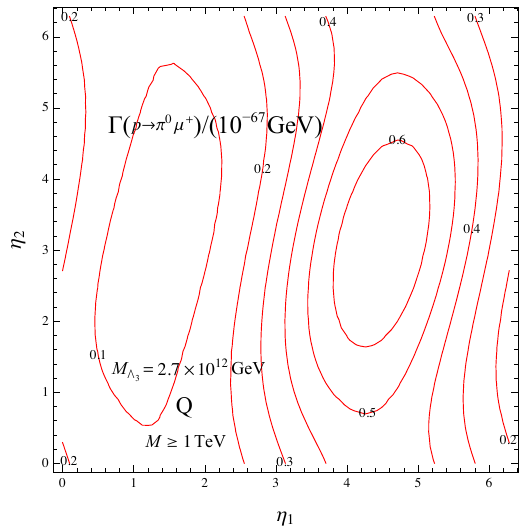}
\includegraphics[width=0.49\textwidth]{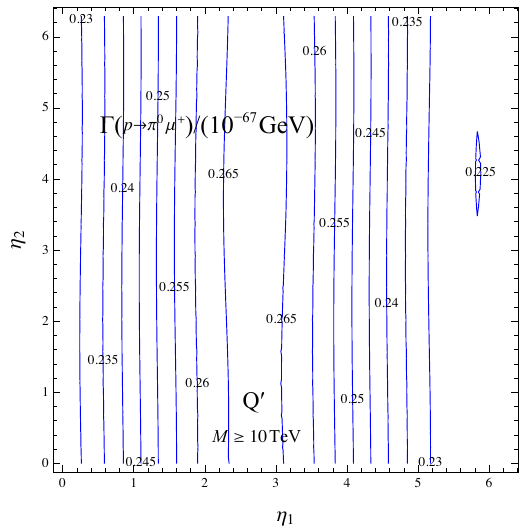}
\includegraphics[width=0.49\textwidth]{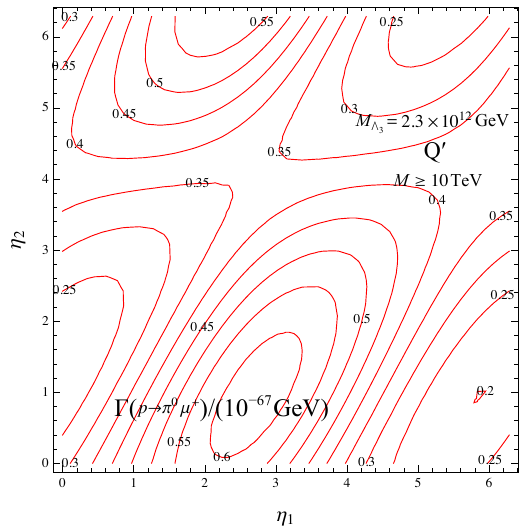}
\caption{\label{fig:eta1eta2} Contours of constant $\Gamma(p \to \pi^0 \mu^+)$ for the gauge boson (blue) and scalar leptoquark (red) mediation for points Q and Q$^\prime$, as indicated.}
\end{figure}

We have already stipulated that variation of $\eta_2$ does not affect gauge mediated proton decay signatures at all. To that end, we present in Fig.~\ref{fig:eta1eta2} contours of constant decay width for $p \to \pi^0 \mu^+$ for both the gauge (blue) and scalar leptoquark (red) mediation in $\eta_1$-$\eta_2$ plane for points Q (upper panels) and Q$^\prime$ (lower panels). It is clear from Fig.~\ref{fig:eta1eta2} that the impact of $\eta_2$ on gauge boson mediated proton decay signatures is marginal, at best. 

Side by side comparison of two types of proton decay mediation, as shown in Fig.~\ref{fig:GSsignatures}, leaves us with the following conclusions. If this model is realized in nature, and if proton is observed to decay to $\pi^0$ and $e^+$, the decay is certainly due to the gauge boson exchange. If the dominant mode is indeed the gauge mediated proton decay, the next channel that is closest to being detected is $p \to \eta^0 e^+$, even though it is predicted to be outside of the projected sensitivity after a ten-year period of data taking at Hyper-K. All other gauge mediated proton decay signatures exceed any reasonable improvement in proton decay bounds to allow for potential discovery in decades to come.

If, on the other hand, proton is observed to decay to $K^+$ and $\overline{\nu}$, the decay is mediated by scalar leptoquark. The channel that is then closest to the ten-year sensitivity projection is $p\to\pi^0\mu^+$. All other channels have predicted lifetimes that vastly exceed what can be probed with current detector technologies.

If both $p\to\pi^0 e^+$ and $p\to K^+\overline{\nu}$ are observed, we would have scenario where both gauge bosons and scalar leptoquark simultaneously contribute towards proton decay signatures. It would then be possible to have an enhancement of the $p\to\pi^0\mu^+$ decay width, as can be seen from Figs.~\ref{fig:GSsignatures} and~\ref{fig:eta1eta2}, if the interference between the gauge and scalar leptoquark contributions at the amplitude level is constructive. Except for potential observation of $p \to \eta^0 e^+$, the remaining four channels will be experimentally inaccessible for all practical purposes within this particular scenario.

The very last points in the parameter space of $M \geq 1$\,TeV and $M \geq 10$\,TeV scenarios in Fig.~\ref{fig:parspace} that could be, at least in principle, probed by proton decay experiments are points A and A$^\prime$, respectively, with common coordinates $(M_{\Phi_1},M_{\Sigma_1})=(10^{12.8}\,\mathrm{GeV},10^{12.9}\,\mathrm{GeV})$. The model yields $\tau_p (p\to\pi^0 e^+) = 1.1 \times 10^{36}$\,years and $\tau_p (p\to\pi^0 e^+) = 1.5 \times 10^{35}$\,years at points A and A$^\prime$, respectively. $M$ is thus inversely proportional to predicted lifetime $\tau_p (p\to\pi^0 e^+)$ at point $(M_{\Phi_1},M_{\Sigma_1})=(10^{12.8}\,\mathrm{GeV},10^{12.9}\,\mathrm{GeV})$. We can accordingly infer that the current experimental limit on $\tau_p (p\to\pi^0 e^+)$ implies that $M \leq 65$\,TeV, where $M$ can be identified, for all practical purposes, with geometric mean of the masses of $\Phi_3$, $\Phi_6$, $\phi_1$, and $\phi_8$. Viable parameter space of our model is nicely bounded from one side by experimental limit on $p \to \pi^0 e^+$ and from the other side by neutrino mass scale perturbativity requirement. The $M = 65$\,TeV scenario is where these two bounds meet.

We can also place an accurate limit on $M_{\Lambda_3}$ within $M \geq 1$\,TeV and $M \geq 10$\,TeV scenarios for any lattice point in $M_{\Phi_1}$-$M_{\Sigma_1}$ plane with the data on $p\to K^+\overline{\nu}$. For example, we find that $M_{\Lambda_3} \geq 1.8 \times 10^{12}$\,GeV, $M_{\Lambda_3} \geq 1.9 \times 10^{12}$\,GeV, $M_{\Lambda_3} \geq 1.8 \times 10^{12}$\,GeV, and $M_{\Lambda_3} \geq 1.7 \times 10^{12}$\,GeV for points Q, R, Q$^\prime$, R$^\prime$, respectively.

Our predictions, up to this point, have been based on the one-loop level gauge coupling unification analysis. We can, in principle, perform the same analysis at the two-loop level. To that end we present the two-loop level prediction for the unification scale at point A with coordinates $(M_{\Phi_1},M_{\Sigma_1})=(10^{12.8}\,\mathrm{GeV},10^{12.9}\,\mathrm{GeV})$ for the $M \geq 1$\,TeV scenario presented in the upper panel of Fig.~\ref{fig:parspace}. The one-loop level analysis at point A yields $M_{\mathrm{GUT}}=1.97 \times 10^{16}$\,GeV and $\alpha_\mathrm{GUT}=0.0457$, whereas the two-loop level maximization of $M_\mathrm{GUT}$ at the same point yields $M_{\mathrm{GUT}}=5.94 \times 10^{16}$\,GeV and $\alpha_\mathrm{GUT}=0.0610$. The mass spectrum that generates this two-loop level unification scenario is 
$M_{\Phi_3}=M_{\Phi_6}=M_{\phi_1}=M_{\phi_8}=1$\,TeV, $M_{\Sigma_6}=4.75 \times 10^{11}$\,GeV, $M_{\Delta_3}= 10^{12}$\,GeV, $M_{\Sigma_3}=3.97 \times 10^{12}$\,GeV, $M_{\Phi_1}=M_{\Phi_{10}}=10^{12.8}$\,GeV, and $M_{\Sigma_1}=10^{12.9}$\,GeV, where we use the two-loop $\beta$-function coefficients given in Appendix~\ref{C}. We accordingly note that the ratio $\alpha_\mathrm{GUT}^2/M_\mathrm{GUT}^4$ at point A changes by a multiplicative factor of 46 when we go from the one-loop to two-loop level result. If we assume that the short-distance coefficients will be enhanced by a factor of 1.1 due to increase of the gauge coupling value in the case of the two-loop level analysis, we can see that the proton decay lifetimes will be scaled by, at most, a factor of 40 for point A. What is important, though, is that the pattern of proton decay signatures shown in Fig.~\ref{fig:GSsignatures} will not be changed if one goes from the one-loop to two-loop level analysis.

Final comments are in order.
\vspace{2.5pt} \\  
($a$) We find an additional multiplicative factor of 4 in denominator of $m_0$ parameter with respect to our previous works~\cite{Dorsner:2019vgf,Dorsner:2021qwg}. This has accordingly reduced available parameter space of the model.
%%%%%%%%%%%%%%%%%%%%%%%%%%%%%%%%%%%
\vspace{2.5pt} \\  
($b$) The matrix elements that we use carry around 10\,\% uncertainty, each, that we do not include in our study of proton decay signatures. These errors would translate to an additional 20\,\% uncertainty for all eight proton decay widths. Again, the uncertainties presented in Table~\ref{tab:per_cent} capture flavor dependence only.
%%%%%%%%%%%%%%%%%%%%%%%%%%%%%%%%%%%
\vspace{2.5pt} \\  
($c$) We remark that, in principle, the fermionic submultiplet $\Sigma_3$ does not need to be almost perfectly degenerate in mass with $\Sigma_6$ and $\Sigma_1$ if one does not require $M_{\Sigma_3},M_{\Sigma_6},M_{\Sigma_1}>0$. This might be problematic for our study since, as shown in Ref.~\cite{Antusch:2023mqe}, a lower mass of $\Sigma_3$, by itself, can significantly increase the unification scale. One might then expect that taking its mass close to the TeV scale would drastically increase $M_\mathrm{GUT}$. This, however, does not happen in our model due to the presence of additional colored light states. For example, if one sets  $\Sigma_3$ to be very light, the maximal unification scale $M_\mathrm{GUT}$ is only a factor of 1.5 larger than what we show in Fig.~\ref{fig:parspace}. Setting $\Sigma_3$ light, however, opens up another issue. To explain it, let us, for example, focus on the third-generation mass. We get, from Eq.~\eqref{massd}, that $m_b\sim m_\tau+\Delta m$, where $\Delta m\sim (5/4) (174\,\mathrm{GeV}) (Y^cY^a) (v_{24}/M_{\Sigma_3})$. To reproduce correct bottom mass, using the data from Table~\ref{tab:input}, one requires $\Delta m\sim 1$\,GeV. Assuming that $\Sigma_3$ state indeed resides at the TeV scale and $v_{24}\sim 2\times 10^{16}$\,GeV, we obtain $Y^c\sim 10^{-17}$, where $Y^a\sim 1$ is used to maximize the neutrino mass scale. We present, in this manuscript, scenario with mass-degenerate $\Sigma_1$, $\Sigma_6$, and $\Sigma_3$ states, which dictates $Y^c$ coupling to be of the order of the usual electron Yukawa coupling and leave thorough discussion of the light $\Sigma_3$ scenario for potential future publication.
%%%%%%%%%%%%%%%%%%%%%%%%%%%%%%%%%%%
\vspace{2.5pt} \\  
($d$) We find that the lower bound on $M_{\Lambda_3}$ is around $10^{12}$\,GeV that somewhat exceeds a value of $3 \times 10^{11}$\,GeV~\cite{Dorsner:2012uz} found in a simple non-renormalizable $SU(5)$ setting. We attribute this change to ($i$) use of short-distance coefficients $A_{SL}$ and $A_{SR}$, ($ii$) implementation of matrix elements instead of more traditional parameters $F$ and $D$ extracted from form factors in semileptonic hyperon decays and nucleon axial charge~\cite{Claudson:1981gh,Aoki:2008ku}, and ($iii$) proper derivation of scalar leptoquark couplings in $SU(5)$~\cite{Dorsner:2012nq}.

%%%%%%%%%%%%%%%%%%%%%%%%%%%%%%%%%%%%%%%%%%%%%%%
%%%%%%%%%%%%%%%%%%%%%%%%%%%%%%%%%%%%%%%%%%%%%%%
\section{Conclusion}
\label{SEC:gcu}

There are only two possible types of mediators of proton decay within the $SU(5)$ model in question. The anticipated experimental signal of these decay processes can, hence, originate solely from exchange of two gauge bosons, or entirely from a single scalar leptoquark exchange, or from combination of the two. In the following, we highlight the most notable features of the proton decay signals of our model.
\begin{itemize}
\item Our analysis stipulates that we can conclude, with certainty, that if a proton is experimentally observed to decay to $\pi^0$ and $e^+$, the decay is mediated by the gauge bosons. If gauge mediation indeed dominates, the only other channel that might potentially be observed is $p \to \eta^0 e^+$, even though it is predicted to be outside of the projected sensitivity after a ten-year period of data taking at Hyper-K.

\item If, on the other hand, proton is observed to decay to $K^+$ and $\overline{\nu}$, the decay is certainly mediated by a scalar leptoquark. One might then hope to observe $p\to\pi^0\mu^+$, even though it is also predicted to be outside of the projected sensitivity after a ten-year period of data taking.

\item If both $p\to\pi^0 e^+$ and $p\to K^+\overline{\nu}$ are observed, the decay width for the process $p\to\pi^0\mu^+$ might be enhanced through fortuitous interference between the gauge and scalar leptoquark contributions at the amplitude level.

\item The model thus predicts that $p \to \pi^+ \overline{\nu}$, $p \to \eta^0 \mu^+$, $p \to K^0 e^+$, and $p \to K^0 \mu^+$ will be experimentally inaccessible in decades to come regardless of the type of proton decay mediation dominance.

\end{itemize}

Other prominent features of the model are as follows. The lightest neutrino is massless and the neutrino mass hierarchy is of the normal ordering. There are four scalar multiplets, i.e., $\Phi_3, \Phi_6 \in 35_H$ and $\phi_1, \phi_8 \in 24_H$, that need to be light if $M_\mathrm{GUT}$ is to attain maximal possible value, where the geometric mean of the masses of these fields can be identified with parameter $M$ for all practical purposes. Current experimental limit on $\tau_p (p\to\pi^0 e^+)$ implies that $M \leq 65$\,TeV, where $M$ and $\tau_p (p\to\pi^0 e^+)$ are inversely proportional. Any future experimental improvement on $\tau_p (p\to\pi^0 e^+)$ will thus lower allowed value of $M$ and accordingly help to further reduce available parameter space of the model.

%%%%%%%%%%%%%%%%%%%%%%%%%%%%%%%%%%%%%%%%%%%%%%%
%%%%%%%%%%%%%%%%%%%%%%%%%%%%%%%%%%%%%%%%%%%%%%%
\section*{Acknowledgments}
I.D.\ acknowledges the financial support from the Slovenian Research Agency (research core funding No.\ P1-0035). S.F.\ acknowledges the financial support from the Slovenian Research Agency (research core funding No.\ P1-0035 and N1-0321).  
S.S.\ would like to thank Kevin Hinze for discussion.

\begin{appendices}
\renewcommand\thesection{\arabic{section}}
\renewcommand\thesubsection{\thesection.\arabic{subsection}}
\renewcommand\thefigure{\arabic{figure}}
\renewcommand\thetable{\arabic{table}}

%%%%%%%%%%%%%%%%%%%%%%%%%%%%%%%%%%%%%%%%
\section{Gauge Mediated Proton Decay Widths}\label{A}

We show below explicit expressions for all eight two-body proton decay widths due to the gauge boson exchange for completeness of exposition. The gauge bosons in question are $X_\mu^{+ 4/3},Y_\mu^{+ 1/3} \in (\overline{3},2,+5/6)$ with a common mass $M_\mathrm{GUT}$.
\begin{equation}
\begin{split}
    \Gamma (p\rightarrow \pi^0 e^+_{\beta})
    &=\frac{(m_p^2-m^2_{\pi})^2}{m^3_p} \frac{\pi}{2} A_{L}^2 \frac{\alpha_{\mathrm{GUT}}^2}{M_\mathrm{GUT}^4}\\
    & \times \Bigr\{ \left|A_{SR} \langle \pi^0 | (ud)_R u_L | p \rangle c(e_{\beta}, d^C)\right|^2+\left|A_{SL} \langle\pi^0|(ud)_L u_L|p\rangle c(e^C_{\beta}, d)\right|^2 \Bigr\}\;.
\end{split}
\end{equation}

\begin{equation}
    \Gamma (p\rightarrow \pi^+ \bar{\nu})
    =\frac{(m_p^2-m^2_{\pi})^2}{m^3_p} \frac{\pi}{2} A_{L}^2 \frac{\alpha_{\mathrm{GUT}}^2}{M_\mathrm{GUT}^4} 
     A_{SR}^2 |\langle\pi^+|(ud)_R d_L|p\rangle|^2 \sum_{i=1}^3 |c(\nu_i , d, d^C)|^2\;.
\end{equation}

\begin{equation}
\begin{split}
    \Gamma (p\rightarrow \eta e^+_{\beta})
    =&\frac{(m_p^2-m^2_{\eta})^2}{m^3_p} \frac{\pi}{2} A_{L}^2 \frac{\alpha_{\mathrm{GUT}}^2}{M_\mathrm{GUT}^4}\\
    &\times \Bigr\{ \left|A_{SR}\langle \eta |(ud)_R u_L|p\rangle c(e_{\beta}, d^C)\right|^2 +\left|A_{SL}\langle \eta |(ud)_L u_L|p \rangle c(e^C_{\beta}, d)\right|^2 \Bigr\}\;.
\end{split}
\end{equation}

\begin{equation}
\begin{split}
    \Gamma (p\rightarrow K^0 e^+_{\beta})
    &=\frac{(m_p^2-m^2_{K})^2}{m^3_p} \frac{\pi}{2} A_{L}^2 \frac{\alpha_{\mathrm{GUT}}^2}{M_\mathrm{GUT}^4}\\
    &\times\Bigr\{ \left|A_{SR}\langle K^0 |(us)_R u_L|p\rangle c(e_{\beta}, s^C)\right|^2+\left|A_{SL}\langle K^0 |(us)_L u_L|p\rangle c(e^C_{\beta}, s)\right|^2 \Bigr\}\;.
\end{split}
\end{equation}

\begin{equation}
\begin{split}
    \Gamma (p\rightarrow K^+ \bar{\nu})
    &=\frac{(m_p^2-m^2_{K})^2}{m^3_p} \frac{\pi}{2} A_{L}^2 \frac{\alpha_{\mathrm{GUT}}^2}{M_\mathrm{GUT}^4}\\
    &\times \Bigr\{ A^2_{SR} \left| \langle K^+ |(us)_R d_L| p \rangle \right|^2 \sum_{i=1}^3 \left| c(\nu_i , d, s^C)\right|^2  \\
    &+ A^2_{SL} \left| \langle K^+ |(ud)_R s_L| p \rangle \right|^2 \sum_{i=1}^3 \left| c(\nu_i , s, d^C)\right|^2  \Bigr\}\;.
\end{split}  
\end{equation}

We use $m_p = 0.9383$\,GeV, $m_\pi = 0.134$\,GeV, $m_\eta = 0.548$\,GeV, and $m_K = 0.493677$\,GeV to generate all our numerical results. The central values of relevant matrix elements~\cite{Aoki:2017puj} are specified in Table~\ref{tab:matrix_elements}.
\begin{table}[h]
\centering
\begin{tabular}{|c|c|}
\hline
Matrix element &  Form factor (GeV$^2$) \\ \hline\hline

$\langle \pi^0| (ud)_R u_L |p \rangle$&  $-0.131$\\
\hline  
$\langle \pi^0| (ud)_L u_L |p \rangle$&  $0.134$\\
\hline\hline  
%%%%%%%%%%%%%%%%%%%%%%%%%%%
$\langle \pi^+| (du)_R d_L |p \rangle$&  $-0.186$\\
\hline 
$\langle \pi^+| (du)_L d_L |p \rangle$&  $0.189$\\
\hline\hline 
%%%%%%%%%%%%%%%%%%%%%%%%%%%
$\langle K^0| (us)_R u_L |p \rangle$&  $0.103$\\
\hline 
$\langle K^0| (us)_L u_L |p \rangle$&  $0.057$\\
\hline\hline  
%%%%%%%%%%%%%%%%%%%%%%%%%%%
$\langle K^+| (us)_R d_L |p \rangle$&  $-0.049$\\
\hline 
$\langle K^+| (ud)_R s_L |p \rangle$&  $-0.134$\\
\hline  
$\langle K^+| (us)_L d_L |p \rangle$&  $0.041$\\
\hline 
$\langle K^+| (ud)_L s_L |p \rangle$&  $0.139$\\
\hline\hline 
%%%%%%%%%%%%%%%%%%%%%%%%%%%
$\langle \eta | (ud)_R u_L |p \rangle$&  $0.006$\\
\hline 
$\langle \eta | (ud)_L u_L |p \rangle$&  $0.113$\\
\hline 

\end{tabular}
\caption{Central values of matrix elements entering the proton decay computation at 2\,GeV scale~\cite{Aoki:2017puj}. For uncertainties associated with each of the elements, see Ref.~\cite{Aoki:2017puj}. }
\label{tab:matrix_elements} 
\end{table}

%%%%%%%%%%%%%%%%%%%%%%%%%%%%%%%%%%%%%%%%
\section{Scalar Mediated Proton Decay Widths}\label{B}

We show below explicit expressions for all eight two-body proton decay widths due to the exchange of a scalar leptoquark $\Lambda_3 (3,1,-1/3)$ of mass $M_{\Lambda_3}$.

\begin{align}
    \Gamma (p\rightarrow \pi^0 e^+_\beta ) 
    &= \frac{(m^2_p-m^2_{\pi})^2}{32 \pi m_p^3 M_{\Lambda_3}^4}A^2_L \nonumber \\
    & \times \Biggl\{ \Bigl| A_{SR} a(d^C, e_\beta)   \langle \pi^0 |(du)_R u_L|p \rangle + A_{SL} a(d, e_\beta)  \langle \pi^0 |(du)_L u_L|p \rangle \Bigr|^2  \nonumber \\
    & +\Bigl| A_{SL} a(d, e^C_\beta)   \langle \pi^0 |(du)_L u_L|p \rangle + A_{SR} a(d^C, e^C_\beta) \langle \pi^0 |(du)_R u_L|p \rangle \Bigr|^2\Biggr\}  \;.
\end{align}

\begin{align}
    \Gamma (p\rightarrow \pi^+ \bar{\nu})
    &=\frac{(m_p^2-m^2_{\pi})^2}{32\pi m^3_p M_{\Lambda_3}^4} A_L^2 \nonumber \\
    &\times \sum_{i=1}^3\Bigl|A_{SR} a(d, d^C,\nu_i) \langle \pi^+ |(du)_R d_L|p \rangle + A_{SL} a(d, d,\nu_i) \langle \pi^+ |(du)_L d_L|p\rangle \Bigr|^2  \;.
\end{align}

\begin{align}
    \Gamma (p\rightarrow \eta^0 e^+_\beta)
    & =  \frac{(m_p^2-m^2_{\eta})^2}{32\pi m^3_p M_{\Lambda_3}^4}A_L^2  \nonumber \\
    &\times \Bigg\{ \Bigl| A_{SL} a(d, e_\beta) \langle \eta |(ud)_L u_L|p\rangle + A_{SR} a(d^C, e_\beta) \langle \eta |(ud)_R u_L|p\rangle\Bigr|^2\nonumber\\
    & +\Bigl| A_{SL} a(d^C, e^C_\beta) \langle \eta |(ud)_L u_L|p\rangle + A_{SR} a(d, e^C_\beta) \langle \eta |(ud)_R u_L|p\rangle\Bigr|^2\Biggr\} \;.
\end{align}

\begin{align}
    \Gamma (p\rightarrow K^0 e^+_\beta) & =   \frac{(m_p^2-m^2_{K})^2}{64\pi m^3_p M_{\Lambda_3}^4}A_L^2 \nonumber\\
    &\times \Biggl\{\Bigr|A_{SR} a(s^C,e_\beta) \langle K^0 |(us)_R u_L|p\rangle+ A_{SL} a(s,e_\beta) \langle K^0 |(us)_L u_L|p\rangle\nonumber\\
    &  -A_{SR} a(s,e^C_\beta)\langle K^0 |(us)_R u_L|p\rangle-A_{SL} a(s^C,e^C_\beta)\langle K^0 |(us)_L u_L|p\rangle\Bigr|^2\nonumber\\
    &  +\Bigr| A_{SR} a(s^C,e_\beta) \langle K^0 |(us)_R u_L|p\rangle+ A_{SL} a(s,e_\beta) \langle K^0 |(us)_L u_L|p\rangle\nonumber\\
    & +A_{SR} a(s,e^C_\beta) \langle K^0 |(us)_R u_L|p\rangle+A_{SL} a(s,e^C_\beta)\langle K^0 |(us)_L u_L|p\rangle\Bigr|^2\Biggr\}  \;.
\end{align}

\begin{align}
    \Gamma (p\rightarrow K^+ \bar{\nu})
    & =\frac{(m_p^2-m^2_{K})^2}{32\pi m^3_p M_{\Lambda_3}^4} A_L^2 \nonumber \\
    &\times \sum_{i=1}^3\left|A_{SL} \left(a(s, d,\nu_i) \langle K^+ |(us)_L d_L|p \rangle + a(d, s,\nu_i) \langle K^+ |(ud)_L s_L|p\rangle\right) \right. \nonumber\\
    &\left. + A_{SR} \left(a(d, s^C,\nu_i) \langle K^+ |(us)_R d_L|p \rangle + a(s, d^C,\nu_i) \langle K^+ |(ud)_R s_L|p\rangle\right) \right|^2 \;.
\end{align}

%%%%%%%%%%%%%%%%%%%%%%%%%%%%%%%%%%%%%%%%
\section{Renormalization Group $\beta$-Function Coefficients}\label{C}
The one-loop coefficients of the renormalization group $\beta$-functions relevant for our model are provided in Table~\ref{table:fields}. In this appendix, following Ref.~\cite{Jones:1981we}, we derive the relevant two-loop coefficients. The two-loop coefficients for the SM read
\begin{align}
b^{\text{SM}}_{ik}=
\left(
\begin{array}{ccc}
 \frac{199}{50} & \frac{27}{10} & \frac{44}{5} \\
 \frac{9}{10} & \frac{35}{6} & 12 \\
 \frac{11}{10} & \frac{9}{2} & -26 \\
\end{array}
\right),    
\end{align}
while the contributions from the new physics states are
\begin{align}
&\Delta b^{ \Lambda_3 }_{ik}= 
\left(
\begin{array}{ccc}
 \frac{4}{75} & 0 & \frac{16}{15} \\
 0 & 0 & 0 \\
 \frac{2}{15} & 0 & \frac{11}{3} \\
\end{array}
\right),\;
%%%%%%%%%%%%%%%%%%%%%%%%%%%%%%%%%
\Delta b^{ \phi_1 }_{ik}= \left(
\begin{array}{ccc}
 0 & 0 & 0 \\
 0 & \frac{28}{3} & 0 \\
 0 & 0 & 0 \\
\end{array}
\right),\;
%%%%%%%%%%%%%%%%%%%%%%%%%%%%%%%%%
\Delta b^{ \phi_8 }_{ik}= \left(
\begin{array}{ccc}
 0 & 0 & 0 \\
 0 & 0 & 0 \\
 0 & 0 & 21 \\
\end{array}
\right),\;
\\%%%%%%%%%%%%%%%%%%%%%%%%%%%%%%%%%
&\Delta b^{ \Phi_1 }_{ik}= \left(
\begin{array}{ccc}
 \frac{729}{25} & 81 & 0 \\
 27 & \frac{245}{3} & 0 \\
 0 & 0 & 0 \\
\end{array}
\right),\;
%%%%%%%%%%%%%%%%%%%%%%%%%%%%%%%%%
\Delta b^{ \Phi_3 }_{ik}= \left(
\begin{array}{ccc}
 \frac{64}{25} & \frac{96}{5} & \frac{64}{5} \\
 \frac{32}{5} & 56 & 32 \\
 \frac{8}{5} & 12 & 11 \\
\end{array}
\right),\;
\\%%%%%%%%%%%%%%%%%%%%%%%%%%%%%%%%%
&\Delta b^{ \Phi_6 }_{ik}= \left(
\begin{array}{ccc}
 \frac{1}{75} & \frac{3}{5} & \frac{8}{3} \\
 \frac{1}{5} & 13 & 40 \\
 \frac{1}{3} & 15 & \frac{230}{3} \\
\end{array}
\right),\;
%%%%%%%%%%%%%%%%%%%%%%%%%%%%%%%%%
\Delta b^{ \Phi_{10} }_{ik}= \left(
\begin{array}{ccc}
 \frac{72}{5} & 0 & 144 \\
 0 & 0 & 0 \\
 18 & 0 & 195 \\
\end{array}
\right),\;
\\%%%%%%%%%%%%%%%%%%%%%%%%%%%%%%%%%
&\Delta b^{ \Sigma_1 }_{ik}= \left(
\begin{array}{ccc}
 \frac{54}{25} & \frac{36}{5} & 0 \\
 \frac{12}{5} & \frac{64}{3} & 0 \\
 0 & 0 & 0 \\
\end{array}
\right),\;
%%%%%%%%%%%%%%%%%%%%%%%%%%%%%%%%%
\Delta b^{ \Sigma_3 }_{ik}= \left(
\begin{array}{ccc}
 \frac{1}{300} & \frac{3}{20} & \frac{4}{15} \\
 \frac{1}{20} & \frac{49}{4} & 4 \\
 \frac{1}{30} & \frac{3}{2} & \frac{38}{3} \\
\end{array}
\right),\;
%%%%%%%%%%%%%%%%%%%%%%%%%%%%%%%%%
\Delta b^{ \Sigma_6 }_{ik}= \left(
\begin{array}{ccc}
 \frac{64}{75} & 0 & \frac{32}{3} \\
 0 & 0 & 0 \\
 \frac{4}{3} & 0 & \frac{125}{3} \\
\end{array}
\right).\;
%%%%%%%%%%%%%%%%%%%%%%%%%%%%%%%%%
\end{align}
Note that multiplets in $\overline \Sigma$ have the same two-loop coefficients as multiplets in $\Sigma$.

\end{appendices}

\bibliographystyle{style}
\bibliography{references}
%%%%%%%%%%%%%%%%%%%%%%%%%%%
\end{document}